\documentclass[aps,twocolumn,showpacs,superscriptaddress,amssymb,floatfix,nofootinbib]{revtex4-1}

\usepackage[english]{babel}
\usepackage{graphicx}
\usepackage{dcolumn}
\usepackage{bm}

\usepackage{xfrac}
\usepackage{physics}
\usepackage{hyperref}
\usepackage{cleveref}
\usepackage{xcolor}
\usepackage{amsmath}

\usepackage{placeins}
\usepackage{tikz}
\usepackage{calc}

\usepackage{verbatim}   

\crefname{subsection}{sec.}{secs.}
\crefname{figure}{fig.}{figs.}
\crefname{equation}{eq.}{eqs.}

\newif\ifshowtikz
\showtikztrue

\let\oldtikzpicture\tikzpicture
\let\oldendtikzpicture\endtikzpicture

\renewenvironment{tikzpicture}{
    \ifshowtikz\expandafter\oldtikzpicture
    \else\comment
    \fi
}{
    \ifshowtikz\oldendtikzpicture
    \else\endcomment
    \fi
}

\newdimen\XCoord
\newdimen\YCoord
\newcommand*{\ExtractCoordinate}[1]{\path (#1); \pgfgetlastxy{\XCoord}{\YCoord};}

\usetikzlibrary{shapes, shapes.geometric,positioning, calc, decorations.pathreplacing, decorations.markings, decorations.pathmorphing, backgrounds, patterns,  fadings, external}
\usepackage{pgffor}

\definecolor{mycolor}{RGB}{255,51,76}
\definecolor{grey}{RGB}{85,98,112}
\definecolor{blue}{RGB}{78,205,196}
\definecolor{yellow}{RGB}{199,244,100}
\definecolor{pink}{RGB}{255,107,107}
\definecolor{red}{RGB}{196,77,88}
\definecolor{green}{RGB}{25,221,137}

\definecolor{purple}{RGB}{168,34,107}

\definecolor{green}{RGB}{150.6509803921568,181.2745098039216,116.0156862745098}
\definecolor{blue}{RGB}{106.7647058823529,193.2705882352941,177.9176470588237}
\definecolor{red}{RGB}{156.862745098039,53.6823529411763,46.5921568627451}
\definecolor{grey}{RGB}{113.798869445446,142.550440681836,145.718104486297}
\definecolor{brown}{RGB}{202.074509803922,149.117647058824,104.192156862745}
\definecolor{pink}{RGB}{214.729411764706,127.058823529412,138.603921568627}

\makeatletter
\begin{document}

\title{Excitations with projected entangled pair states using the corner transfer matrix method}

\author{Boris \surname{Ponsioen}} \affiliation{Institute for Theoretical Physics Amsterdam and Delta Institute for Theoretical Physics, University of Amsterdam, Science Park 904, 1098 XH Amsterdam, The Netherlands}
\author{Philippe \surname{Corboz}}  \affiliation{Institute for Theoretical Physics Amsterdam and Delta Institute for Theoretical Physics, University of Amsterdam, Science Park 904, 1098 XH Amsterdam, The Netherlands}

\date{\today}

\begin{abstract}
  We present an extension of a framework for simulating single quasiparticle or collective excitations on top of strongly correlated quantum many-body ground states using infinite projected entangled pair states, a tensor network ansatz for two-dimensional wave functions in the thermodynamic limit.
  Our approach performs a systematic summation of locally perturbed states in order to obtain excited eigenstates localized in momentum space, using the corner transfer matrix method, and generalizes the framework to arbitrary unit cell sizes, the implementation of global Abelian symmetries and fermionic systems.
  Results for several test cases are presented, including the transverse Ising model, the spin-$\sfrac{1}{2}$ Heisenberg model and a free fermionic model, to demonstrate the capability of the method to accurately capture dispersions. We also provide insight into the nature of excitations at the $k=(\pi,0)$ point of the Heisenberg model.
\end{abstract}

\maketitle

\tikzstyle{HB} = [preaction={fill, blue}, pattern=flexible hatch, pattern color=red, hatch distance=0.3cm, hatch thickness=0.05cm, line width=0.05cm, minimum height=0.75cm, minimum width=0.75cm, rounded corners=0.1cm, text=white, draw=blue]

\tikzstyle{N} = [draw=grey, line width=0.05cm, fill=white, minimum height=0.75cm, minimum width=0.75cm, rounded corners=0.1cm, text=grey]
\tikzstyle{H} = [preaction={fill, white}, pattern=flexible hatch, pattern color=red, hatch distance=0.3cm, hatch thickness=0.05cm, draw=red, line width=0.05cm, minimum height=0.75cm, minimum width=0.75cm, rounded corners=0.1cm, text=red, text opacity=0]
\tikzstyle{B} = [draw=blue, line width=0.05cm, fill=blue, minimum height=0.75cm, minimum width=0.75cm, rounded corners=0.1cm, text opacity=1, text=white]

\tikzstyle{tensor} = [draw=grey, line width=0.05cm, fill=white, minimum height=0.4cm, circle, inner sep=0cm]
\tikzstyle{tensorxin} = [draw=grey, line width=0.05cm, fill=white, minimum height=0.4cm, circle, text=red, inner sep=0cm, fill opacity=0, text opacity=1, draw opacity=0]
\tikzstyle{tensorx} = [draw=red, line width=0.05cm, fill=white, minimum height=0.4cm, circle, text=red, inner sep=0cm]
\tikzstyle{tensorb} = [draw=blue, line width=0.05cm, fill=blue, minimum height=0.4cm, circle, text=blue, inner sep=0cm]
\tikzstyle{tensorxb} = [draw=blue, line width=0.05cm, fill=blue, minimum height=0.4cm, circle, text=red, inner sep=0cm]

\tikzstyle{proj} = [regular polygon,regular polygon sides=3, shape border rotate=180, draw=black, fill=black, line width=0.05cm,inner sep=0cm, minimum size=0.45cm, rounded corners=0.05cm, text=green, text opacity=0]
\tikzstyle{Hconnecth} = [draw=red, line width=0.00cm, fill=red, inner sep=0cm, minimum width=1.2cm, minimum height=0.25cm, rounded corners=0.0cm, text=blue, fill opacity=0.7, draw opacity=0]
\tikzstyle{Hconnectv} = [draw=red, line width=0.00cm, fill=red, inner sep=0cm, minimum width=0.25cm, minimum height=1.2cm, rounded corners=0.0cm, text=blue, fill opacity=0.7, draw opacity=0]

\tikzstyle{T} = [minimum width=0.4cm, rounded corners=0.15cm, inner sep=0cm]
\tikzstyle{Th} = [minimum height=0.4cm, rounded corners=0.15cm, inner sep=0cm]
\tikzstyle{Ch} = [minimum width=1.65cm]
\tikzstyle{Cv} = [minimum height=1.65cm]
\tikzstyle{TTh} = [minimum width=1.60cm]
\tikzstyle{TTv} = [minimum height=1.60cm]

\tikzset{
  hatch distance/.store in=\hatchdistance,
  hatch distance=10pt,
  hatch thickness/.store in=\hatchthickness,
  hatch thickness=2pt
}

\tikzset{
  projector/.pic={
    \node[proj, name=proj] at (0.5,-0.75) {};
    \draw[line width=0.1cm] (0,-0.4) to[out=270,in=90] ($(proj)+(-0.05,0.12)$);
    \draw[doubline] (1,-0.4) to[out=270,in=90] ($(proj)+(0.05,0.12)$);
    \draw[line width=0.1cm] ($(proj)+(0,-0.12)$) -- ($(proj)+(0,-0.35)$);
    \node[proj] at (proj) {};
}}
\tikzset{
  projector up/.pic={
    \node[proj, name=proj, shape border rotate=0] at (0.5,0.75) {};
    \draw[line width=0.1cm] (0,0.4) to[out=90,in=270] ($(proj)+(-0.05,-0.12)$);
    \draw[doubline] (1,0.4) to[out=90,in=270] ($(proj)+(0.05,-0.12)$);
    \draw[line width=0.1cm] ($(proj)+(0,0.12)$) -- ($(proj)+(0,0.35)$);
    \node[proj, shape border rotate=0] at (proj) {};
}}
\tikzset{
  projector up high/.pic={
    \node[proj, name=proj, shape border rotate=0] at (0.5,1.75) {};
    \draw[line width=0.1cm] (0,1.4) to[out=90,in=270] ($(proj)+(-0.05,-0.12)$);
    \draw[doubline] (1,1.4) to[out=90,in=270] ($(proj)+(0.05,-0.12)$);
    \draw[line width=0.1cm] ($(proj)+(0,0.12)$) -- ($(proj)+(0,0.35)$);
    \node[proj, shape border rotate=0] at (proj) {};
}}

\makeatletter
\pgfdeclarepatternformonly[\hatchdistance,\hatchthickness]{flexible hatch}
{\pgfqpoint{0pt}{0pt}}
{\pgfqpoint{\hatchdistance}{\hatchdistance}}
{\pgfpoint{\hatchdistance-1pt}{\hatchdistance-1pt}}
{
  \pgfsetcolor{\tikz@pattern@color}
  \pgfsetlinewidth{\hatchthickness}
  \pgfpathmoveto{\pgfqpoint{0pt}{0pt}}
  \pgfpathlineto{\pgfqpoint{\hatchdistance}{\hatchdistance}}
  \pgfusepath{stroke}
}
\makeatother

\tikzset{doubline/.style={
    preaction={
      draw=black,
      double distance between line centers=0.075cm,
      line width=0.04cm,
      shorten >=0pt,
      shorten <=0pt,
    },
    draw=white,
    line width=0.075cm,
    shorten >=-.1pt,
    shorten <=-.1pt,
  }}

 \section{Introduction}
\label{sec:introduction}

At the center of condensed matter physics lies the problem of understanding the behavior of strongly correlated many-body systems.
In low-dimensional models, strong quantum effects lead to a wealth of interesting and at times unexpected phenomena, yet also to extreme challenges in simulations and often make analytical treatments infeasible.
One of the most important concepts in understanding physics in quantum many-body systems is the idea of quasiparticles or collective excitations as being the low-lying excitations on top of a strongly interacting ground state~\cite{landauTheorySuperfluidityHelium1941}.
Examples of such excitations are quasiparticles made of combinations of particles and holes in fermionic models, magnons as quanta of spin waves in quantum magnets or phonons that determine elastic responses in solids~\cite{debyeZurTheorieSpezifischen1912}.

This description is not only useful for analytical methods~\cite{andersonConceptsSolidsLectures1997}, due to its simplicity that stands in sharp contrast to the extremely complicated strongly correlated ground-state wave functions in many systems, but also leads to an intuitive description in the context of tensor networks~\cite{haegemanVariationalMatrixProduct2012,haegemanElementaryExcitationsGapped2013a,haegemanPostmatrixProductState2013,zaunerTransferMatricesExcitations2015,vanderstraetenExcitationsTangentSpace2015,vanderstraetenSimulatingExcitationSpectra2019}.
Matrix product states (MPS)~\cite{ostlundThermodynamicLimitDensity1995a}, a type of tensor network ansatz for (quasi) one dimensional systems, have been widely used to simulate quantum many-body ground states ever since the conception of the DMRG method~\cite{whiteDensityMatrixFormulation1992}.
The quasiparticle concept can also be directly applied to ground states encoded by MPS, resulting in a powerful technique for the study of excitations~\cite{haegemanVariationalMatrixProduct2012,zaunerTransferMatricesExcitations2015,haegemanElementaryExcitationsGapped2013a,haegemanPostmatrixProductState2013}.
In this method, a summation over local perturbations is performed in a systematic way to create an excitation that is localized in momentum space.
It has been successfully applied in many contexts, such as magnons in spin chains, spin and charge excitations in the Hubbard model and even scattering states and topologically non-trivial excitations~\cite{zauner-stauberTopologicalNatureSpinons2018,vandammeRealtimeScatteringInteracting2019}.
An alternative technique to gain insight into the low-lying excitations with MPS is to compute dynamical structure factors, which has been used both in one dimensional systems~\cite{hallbergDensitymatrixAlgorithmCalculation1995,kuhnerDynamicalCorrelationFunctions1999,jeckelmannDynamicalDensitymatrixRenormalizationgroup2002,whiteRealTimeEvolutionUsing2004,holznerChebyshevMatrixProduct2011,noceraSpectralFunctionsDensity2016,bruognoloDynamicStructureFactor2016} as well on two-dimensional semi-infinite cylinders~\cite{zaletelTimeevolvingMatrixProduct2015,gohlkeDynamicsKitaevHeisenbergModel2017,verresenQuantumDynamicsSquarelattice2018}.

Projected entangled pair states (PEPS)~\cite{verstraeteMatrixProductStates2008,murgVariationalStudyHardcore2007,verstraeteRenormalizationAlgorithmsQuantumMany2004} (or tensor product states~\cite{nishinoTwoDimensionalTensorProduct2001,nishioTensorProductVariational2004}) are a natural extension of MPS to higher dimensional lattice systems.
Similarly to MPS, the PEPS ansatz can be applied in the thermodynamic limit, referred to as infinite PEPS (iPEPS).
Recently the quasiparticle-based method for simulating excitations has been extended to iPEPS~\cite{vanderstraetenExcitationsTangentSpace2015,vanderstraetenSimulatingExcitationSpectra2019}, opening up the possibility to study two-dimensional infinite systems.
Since iPEPS cannot be contracted exactly and approximate contraction schemes are needed, the summation over real space perturbations constitutes significant challenges.
First applied in frustration-free Hamiltonian models~\cite{vanderstraetenExcitationsTangentSpace2015}, yielding an accurate value of the gap in the Affleck-Kennedy-Lieb-Tasaki model, the iPEPS excitation ansatz has recently correctly reproduced the behavior of the spin-$\sfrac{1}{2}$ Heisenberg antiferromagnet~\cite{vanderstraetenSimulatingExcitationSpectra2019}, where the energy deviates from conventional linear spin wave theory and the underlying physics is still debated~\cite{dallapiazzaFractionalExcitationsSquarelattice2015,christensenQuantumDynamicsEntanglement2007,powalskiMutuallyAttractingSpin2018,shaoNearlyDeconfinedSpinon2017a,verresenQuantumDynamicsSquarelattice2018}.

In this paper we provide an alternative contraction scheme to power the iPEPS excitation ansatz, based on the corner transfer matrix renormalization group method (CTM) ~\cite{nishinoCornerTransferMatrix1996,nishinoCornerTransferMatrix1997,orusSimulationTwodimensionalQuantum2009} that has been widely used in iPEPS algorithms.
The summations for the momentum superposition are performed in a manner similar to the variational optimization method in Ref.~\cite{corbozVariationalOptimizationInfinite2016} by systematically keeping track of all relevant contributions in a growing system until convergence.
Furthermore, we extend the capabilities of the excitation scheme in multiple directions: (1)~arbitrary unit cell sizes, enabling simulations of states with partially broken translational symmetry; (2)~the ability to enforce global Abelian symmetries, which can be used to restrict excitations to certain symmetry sectors and greatly reduces computational cost and lastly (3)~fermionic systems.

We test our framework on several models, starting with the quantum transverse field Ising model, in which we show that the accuracy of the dispersions for field strengths $h=2.5,3$ is in close agreement to the results in Ref.~\cite{vanderstraetenSimulatingExcitationSpectra2019}.
Additionally, we study the behavior of the second lowest excitation, which consists of two-particle bound states that can also be captured by our approach, as a function of the field strength.
Then we move on to the spin-$\sfrac{1}{2}$ Heisenberg model, with a special focus on the spin wave anomalous point at $k=(\pi,0)$ where we find the dispersion to agree well with various numerical and experimental results, and we provide local real-space visualizations of the excitations.
Finally, we conclude with a free fermionic model with an additional pairing term, to demonstrate that in the gapped phase the dispersions can be computed very accurately with a small number of free parameters, and systematically approach the dispersion in the gapless phase.

\section{Excitations in iPEPS}
\label{sec:excitation_ctm}

\subsection{iPEPS}
\label{sub:ipeps}

In the iPEPS ansatz of the ground-state wave function of a two-dimensional quantum system, the $\order{dD^4}$ variational parameters are contained in order-5 tensors $A$, where $d$ corresponds to the local Hilbert space of a single site in the system, and the \emph{bond dimension} $D$ systematically controls the accuracy of the ansatz.
By exploiting translational invariance, iPEPS can describe states in the thermodynamic limit with a single tensor.
For states with partially broken translational symmetry, where the state is comprised of repetitions of a unit cell, different tensors $A_x$ are used for each site $x=(i,j)$ in the unit cell.

We use a highly precise ground-state optimization algorithm based on the ideas from Ref.~\cite{corbozVariationalOptimizationInfinite2016} to obtain the ground state $A$ tensors, since the accuracy of the ground state greatly impacts the accuracy of the excited states.

In practice the contraction of the network cannot be performed exactly and approximate contraction schemes are necessary.
Here we build upon the corner transfer matrix renormalization group (CTM)~\cite{nishinoCornerTransferMatrix1996,orusSimulationTwodimensionalQuantum2009} method, which approximates the contraction of iPEPS networks by iteratively growing the lattice around a center site and truncating the tensors that contain the environment down to the most relevant subspace.
Once the CTM algorithm has converged, the environment approximates the full infinite network and properties of the iPEPS in the thermodynamic limit can be computed.

\subsection{Excitation Ansatz}
\label{sub:excitation_ansatz}

\begin{figure}[htpb]
  \begin{tikzpicture}[scale=0.8, transform shape]
    \begin{scope}[local bounding box=scope1, shift={(0,0)}]
      \begin{scope}[shift={(1.5,-2)}]

        \foreach \x in {0,1,2,3,4}
        {
          \draw[line width=0.07cm] ($(\x,-0.5)$) -- ($(\x,4.5)$);
          \draw[line width=0.07cm] ($(-0.5,\x)$) -- ($(4.5,\x)$);
        }

        \foreach \x in {0,1,2,3,4}
        {
          \foreach \y in {0,1,2,3,4}
          {
            \draw[line width=0.05cm,grey] ($(\x,\y)$) -- ($(\x,\y)+(0.3,-0.3)$);
            \node[tensor] at ($(\x,\y)$) {};
          }
        }
        \node[tensor, draw=blue, fill=blue] at (2,2) {};

        \node[anchor=east] at (-0.5,2) {$\dots$};
        \node[anchor=north] at (2,-0.3) {$\vdots$};
        \node[anchor=south] at (2,4.5) {$\vdots$};
        \node[anchor=west] at (4.5,2) {$\dots$};

      \end{scope}

    \end{scope}
  \end{tikzpicture}
  \caption{Schematic picture of an excitation tensor in the center of an infinite PEPS.}
  \label{fig:peps_exci_center}
\end{figure}
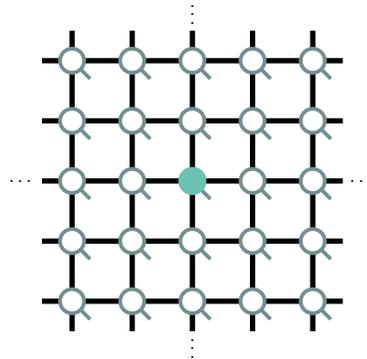

An elementary excitation, consisting of a single quasiparticle localized in momentum space, can be approximately described by a perturbation of an iPEPS ground state~\cite{vanderstraetenExcitationsTangentSpace2015}.
One of the ground-state tensors $A$ is replaced by a different tensor $B$ on a location $x$, which we write as
\begin{equation}
  \ket{\Phi(B)_x}
\end{equation}
and picture diagrammatically in~\Cref{fig:peps_exci_center}.
Then the excited eigenstate with momentum $k$ is obtained by a superposition of such states
\begin{equation}
  \ket{\Phi(B)_k} = \sum_x e^{i k x} \ket{\Phi(B)_x}~.
\end{equation}
The computation of the energy of this state, where we denote a local Hamiltonian term by $\mathcal{H}_{n}$, is given by
\begin{multline}
  \mel{\Phi(B)_k}{\mathcal{H}}{\Phi(B)_k} = \\
  \sum_{x_1,x_2,n} e^{-i k (x_1 - x_2))} \mel{\Phi(B)_{x_1}}{\mathcal{H}_{n}}{\Phi(B)_{x_2}}~.
\end{multline}
This requires a triple infinite sum, which can be reduced to a double infinite sum by exploiting translational invariance of the ground state.

We write the overlap between the ground-state iPEPS $\ket{\Psi}$ and an excited state as $\braket{\Psi}{\Phi(B)_k} \equiv \delta(k) \va{g}^\dagger \va{B}$ with $\va{B}$ the vectorized representation of $B$.
Then we can form a complete basis of vectors $\va{b}^m$ that are orthogonal to $\va{g}$, i.e. $\va{b}^m \in \mathrm{null}\qty(\va{g})$, so that each $\ket{\Phi(b^m)_k}$ is orthogonal to the ground state.
We can then evaluate the matrix elements of the effective norm matrix
\begin{equation}
  \mathbb{N}^{ij}_k = \braket{\Phi(b^i)_k}{\Phi(b^j)_k}~.
\label{eq:norm_matrix}
\end{equation}

More involved is the evaluation of the effective energy matrix elements
\begin{equation}
\mathbb{H}^{ij}_k = \mel{\Phi(b^i)_k}{\mathcal{H}}{\Phi(b^j)_k}
\label{eq:energy_matrix}
\end{equation}
for which we provide a detailed description in~\Cref{sub:excitations_ctm}.

Generally, these matrices are not well conditioned due to the presence of modes with zero norm: in any state with a $B$ tensor of the form
\begin{equation}
  B_x = e^{i k} A_x \cdot M_x - M_x \cdot A_x
\end{equation}
with $M_x$ any $D \times D$ matrix, the terms in the momentum superposition will cancel exactly~\cite{haegemanPostmatrixProductState2013,vanderstraetenExcitationsTangentSpace2015}.
Therefore we use the eigendecomposition of $\mathbb{N} = v \Lambda v^{\dag}$ to compute a reduced basis $P=\tilde{v}$ in which the basis vectors corresponding to eigenvalues close to zero have been removed.
In the reduced basis, we can formulate a generalized eigenvalue problem
\begin{equation}
  P^\dag \mathbb{H}_k P = \omega_k P^\dag \mathbb{N}_k P
  \label{eq:main_eig_problem}
\end{equation}
where $\omega_k$ corresponds to the energy of the eigenmodes which form the excited states.
By performing this procedure for each momentum $k$ the dispersion through the Brillouin zone can be computed.

\subsection{Excitations with CTM}
\label{sub:excitations_ctm}

In Refs.~\cite{vanderstraetenExcitationsTangentSpace2015,vanderstraetenSimulatingExcitationSpectra2019} it was shown that this representation of elementary excitations with iPEPS is able to accurately reproduce the dispersion of several spin models.
The double infinite sum in the computation of the energy was handled by so-called channel environments, which are an extension of the MPS-based contraction scheme that has been used successfully in iPEPS ground-state simulations~\cite{vanderstraetenGradientMethodsVariational2016}.

Here we introduce a different approach, based on the CTM contraction method~\cite{nishinoCornerTransferMatrix1996,orusSimulationTwodimensionalQuantum2009}, for computing the matrix elements of~\Cref{eq:norm_matrix,eq:energy_matrix}.

\subsubsection{Main scheme}
\label{ssub:main_scheme}

The contraction of all tensors in both bra and ket layers around a certain site, here referred to as the \emph{environment} of that site, can be approximated by a set of boundary tensors.
With these tensors, the norm of an iPEPS can be approximated by the following contraction:

\begin{equation}
    \begin{tikzpicture}[scale=1, transform shape]

      \begin{scope}[local bounding box=scope1, shift={(0,0)}]
        \node[anchor=west] at (-1,-0.5) {$\braket{\Psi} =$};

        \begin{scope}[shift={(0,-2)}]
          \draw[doubline] (0,0.5) -- (0,-2.5);
          \draw[doubline] (1,0.5) -- (1,-2.5);
          \draw[doubline] (2,0.5) -- (2,-2.5);

          \draw[doubline] (-0.5,0) -- (2.5,0);
          \draw[doubline] (-0.5,-1) -- (2.5,-1);
          \draw[doubline] (-0.5,-2) -- (2.5,-2);

          \node[tensor] at (0,0) {};
          \node[tensor] at (2,0) {};
          \node[tensor] at (2,-2) {};
          \node[tensor] at (0,-2) {};
          \node[tensor] at (1,0) {};
          \node[tensor] at (2,-1) {};
          \node[tensor] at (1,-2) {};
          \node[tensor] at (0,-1) {};
          \node[tensor] at (1,-1) {};

          \node[] at (3,-1) {$\dots$};
          \node[] at (-1,-1) {$\dots$};
          \node[] at (1,1) {$\vdots$};
          \node[] at (1,-3) {$\vdots$};

          \node[] at (3.5,-1) {$\approx$};
        \end{scope}

        \begin{scope}[shift={(4.5,-2)}]
          \draw[line width=0.1cm] (0,0) -- (0,-2);
          \draw[line width=0.1cm] (0,0) -- (2,0);
          \draw[line width=0.1cm] (0,-2) -- (2,-2);
          \draw[line width=0.1cm] (2,0) -- (2,-2);
          \draw[doubline] (1,0) -- (1,-2);
          \draw[doubline] (0,-1) -- (2,-1);
          \draw[line width=0.1cm] (1,0) -- (1.5,0);

          \node[N] at (0,0) {{\small $C_1$ }};
          \node[N] at (2,0) {{\small $C_2$ }};
          \node[N] at (2,-2) {{\small $C_3$ }};
          \node[N] at (0,-2) {{\small $C_4$ }};
          \node[N, T] at (1,0) {{\small $T_1$ }};
          \node[N, Th] at (2,-1) {{\small $T_2$ }};
          \node[N, T] at (1,-2) {{\small $T_3$ }};
          \node[N, Th] at (0,-1) {{\small $T_4$ }};
          \node[tensor] at (1,-1) {};

        \end{scope}
      \end{scope}

    \end{tikzpicture}
\end{equation}
The grey shapes represent the boundary tensors that approximate the environment of the center site.
These boundary tensors are labelled $C_1 \dots C_4$ for the corner tensors (corner transfer matrices) and $T_1 \dots T_4$ for the half-row and half-column transfer matrices.
The thick lines connecting the boundary tensors represent indices of size $\chi$, referred to as the \emph{boundary bond dimension}, which controls the accuracy of the environment.

In the figures we take a top-down view of the network in which the bra and ket layer tensors are stacked on top of each other for brevity.
The grey circles represent pairs of bra and ket tensors on each site and double lines correspond to the two $D$ indices that connect tensors within each layer, as follows:
\begin{equation}
  \begin{tikzpicture}[scale=1, transform shape]

    \begin{scope}[local bounding box=scope1, shift={(0,0)}]
      \begin{scope}[shift={(0,0)}]
        \draw[doubline] (-0.5,0) -- (0.5,0);
        \draw[doubline] (0,-0.5) -- (0,0.5);

        \node[tensor] at (0,0) {};

        \node[] at (1.15,0) {$=$};
      \end{scope}

    \end{scope}

    \begin{scope}[local bounding box=scope1, shift={(2.5,0.5)}]
      \begin{scope}[shift={(0,0)}]
        \draw[line width=0.075cm,grey] (0,0) -- (0,-1);
        \draw[line width=0.05cm] (-0.5,0) -- (0.5,0);
        \draw[line width=0.05cm] (-0.5,-0.3) -- (0.5,0.3);
        \draw[line width=0.05cm] (-0.5,-1) -- (0.5,-1);
        \draw[line width=0.05cm] (-0.5,-1.3) -- (0.5,-0.7);

        \node[tensor] at (0,0) {};
        \node[tensor] at (0,-1) {};
      \end{scope}

    \end{scope}

  \end{tikzpicture}
\end{equation}

The boundary tensors are computed through an iterative procedure, in which at each step rows and columns of sites are contracted with the boundary tensors of the previous step, once for every direction within each step.
In this section we focus on the contraction of a column of sites into the left side boundary tensors (\emph{left move}); the other directional moves are equivalent up to simple rotations.
Performing the contractions exactly would lead to an exponential growth of the number of elements in the boundary tensors, so an approximation has to be implemented.
In this approximation, the updated boundary tensors are truncated to a given boundary bond dimension $\chi$ by \emph{projectors}~\cite{wangMonteCarloSimulation2011,huangAccurateComputationLowtemperature2012,corbozCompetingStatesModel2014}, which we set to a sufficiently high value such that the finite-$\chi$ error is negligible compared to the finite-$D$ error.
The projectors can be computed in several ways, though they provide the same results in the large $\chi$ limit~\footnote{We compute the projectors based on the ground-state environment, however it remains an open question whether projectors that take into account the perturbed state could provide better results.}.
For example, the left row transfer matrix $T_4$, representing an infinite row of sites extending to the left of the unit cell, is updated by absorbing a new site and then truncated down to a $\chi \times \chi \times D \times D$ tensor in the following way during the left move:
\begin{equation}
  \begin{tikzpicture}[scale=1, transform shape]

    \begin{scope}[local bounding box=scope1, shift={(0,0)}]
      \node[anchor=west] at (-1.5,0) {$T_4'$};

      \begin{scope}
        \draw[line width=0.1cm] (0,0) -- (0,0.5);
        \draw[doubline] (0,0) -- (0.75,0);
        \draw[line width=0.1cm] (0,0) -- (0,-0.5);
        \node[N, Th] at (0,0) {};
        \node[] at (1.25,0) {$=$};
      \end{scope}

      \begin{scope}[shift={(2.25,0)}]
        \draw[line width=0.1cm] (0,0) -- (0,0.4);
        \draw[doubline] (0,0) -- (1,0);
        \draw[line width=0.1cm] (0,0) -- (0,-0.4);

        \draw[doubline] (1,0) -- (1,0.4);
        \draw[doubline] (1,0) -- (1.5,0);
        \draw[doubline] (1,0) -- (1,-0.4);

        \node[N, Th] at (0,0) {};
        \node[tensor] at (1,0) {};

        \pic{projector};
        \pic{projector up};
      \end{scope}
    \end{scope}

  \end{tikzpicture}
\end{equation}
where the black triangular shapes represent the projectors.
Similarly, the corner transfer matrix $C_1$, containing all sites in the upper left corner of the network, is updated as
\begin{equation}
  \begin{tikzpicture}[scale=1, transform shape]
    \begin{scope}[local bounding box=scope1, shift={(0,0)}]
      \node[anchor=west] at (-1.5,0) {$C_1'$};

      \begin{scope}
        \draw[line width=0.1cm] (0,0) -- (0,-.75);
        \draw[line width=0.1cm] (0,0) -- (0.75,0);
        \node[N] at (0,0) {};
      \end{scope}

      \begin{scope}[shift={(1.25,0)}]
        \node[] at (0,0) {$=$};
      \end{scope}

      \begin{scope}[shift={(2.25,0)}]
        \draw[line width=0.1cm] (0,0) -- (0,-.4);
        \draw[line width=0.1cm] (0,0) -- (1,0);
        \draw[doubline] (1,0) -- (1,-0.4);
        \draw[line width=0.1cm] (1,0) -- (1.5,0);
        \node[N] at (0,0) {};
        \node[N, T] at (1,0) {};

        \pic{projector};
      \end{scope}

    \end{scope}
  \end{tikzpicture}
\end{equation}
This procedure is repeated until convergence with respect to the singular values of the corner matrices or expectation values of observables calculated using the environment.

The CTM method, used primarily for measuring local observables in iPEPS ground states, has previously been extended to compute also effective energy environments~\cite{corbozVariationalOptimizationInfinite2016} that can be used to calculate gradients for highly accurate ground-state optimization algorithms.
Such energy environments consist of an infinite summation of Hamiltonian terms around a center unit cell of an iPEPS ground state.

By extending this idea we are able to perform also higher order summations in an iterative manner, in a similar way to a recent application in the context of the Tensor Renormalization Group (TRG) algorithm~\cite{moritaCalculationHigherorderMoments2019} for classical systems.
If we denote a single term in the overlap $\braket{\Psi(A)}{\Phi(B)_k}$ between an excited state and the ground state by a colored tensor in the network, meaning that in the ket layer one ground-state tensor $A$ is swapped for a $B$ tensor, the contraction can be performed using the regular norm environments:
\begin{equation}
  \begin{tikzpicture}[scale=1, transform shape]

    \begin{scope}[local bounding box=scope1, shift={(0,0)}]
      \node[anchor=west] at (-1,-0.5) {$\braket{\Psi}{\Phi(B)_{(0,0)}} =$};

      \begin{scope}[shift={(0,-2)}]
        \draw[doubline] (0,0.5) -- (0,-2.5);
        \draw[doubline] (1,0.5) -- (1,-2.5);
        \draw[doubline] (2,0.5) -- (2,-2.5);

        \draw[doubline] (-0.5,0) -- (2.5,0);
        \draw[doubline] (-0.5,-1) -- (2.5,-1);
        \draw[doubline] (-0.5,-2) -- (2.5,-2);

        \node[tensor] at (0,0) {};
        \node[tensor] at (2,0) {};
        \node[tensor] at (2,-2) {};
        \node[tensor] at (0,-2) {};
        \node[tensor] at (1,0) {};
        \node[tensor] at (2,-1) {};
        \node[tensor] at (1,-2) {};
        \node[tensor] at (0,-1) {};
        \node[tensor, draw=blue, fill=blue] at (1,-1) {};

        \node[] at (3,-1) {$\dots$};
        \node[] at (-1,-1) {$\dots$};
        \node[] at (1,1) {$\vdots$};
        \node[] at (1,-3) {$\vdots$};

        \node[] at (3.5,-1) {$\approx$};
      \end{scope}

      \begin{scope}[shift={(4.5,-2)}]
        \draw[line width=0.1cm] (0,0) -- (0,-2);
        \draw[line width=0.1cm] (0,0) -- (2,0);
        \draw[line width=0.1cm] (0,-2) -- (2,-2);
        \draw[line width=0.1cm] (2,0) -- (2,-2);
        \draw[doubline] (1,0) -- (1,-2);
        \draw[doubline] (0,-1) -- (2,-1);
        \draw[line width=0.1cm] (1,0) -- (1.5,0);

        \node[N] at (0,0) {};
        \node[N] at (2,0) {};
        \node[N] at (2,-2) {};
        \node[N] at (0,-2) {};
        \node[N, T] at (1,0) {};
        \node[N, Th] at (2,-1) {};
        \node[N, T] at (1,-2) {};
        \node[N, Th] at (0,-1) {};
        \node[tensor, draw=blue, fill=blue] at (1,-1) {};

      \end{scope}
    \end{scope}

  \end{tikzpicture}
\end{equation}
where
\begin{equation}
  \begin{tikzpicture}[scale=1, transform shape]

    \begin{scope}[local bounding box=scope1, shift={(0,0)}]
      \begin{scope}[shift={(0,0)}]
        \draw[doubline] (-0.5,0) -- (0.5,0);
        \draw[doubline] (0,-0.5) -- (0,0.5);

        \node[tensorb] at (0,0) {};

        \node[] at (1.15,0) {$=$};
      \end{scope}

    \end{scope}

    \begin{scope}[local bounding box=scope1, shift={(2.5,0.5)}]
      \begin{scope}[shift={(0,0)}]
        \draw[line width=0.075cm,grey] (0,0) -- (0,-1);
        \draw[line width=0.05cm] (-0.5,0) -- (0.5,0);
        \draw[line width=0.05cm] (-0.5,-0.3) -- (0.5,0.3);
        \draw[line width=0.05cm] (-0.5,-1) -- (0.5,-1);
        \draw[line width=0.05cm] (-0.5,-1.3) -- (0.5,-0.7);

        \node[tensorb] at (0,0) {};
        \node[tensor] at (0,-1) {};
      \end{scope}

    \end{scope}

  \end{tikzpicture}
\end{equation}
consists of a pair of a $B$ tensor in the ket layer and an $A^{\dagger}$ tensor in the bra layer. 
Note that we choose the $B$ such that this overlap between a locally perturbed state and the ground state is always zero.

By placing another tensor $B'^\dagger$ in the center of the bra layer, we can compute an overlap of the form $\braket{\Phi(B')_{(0,0)}}{\Phi(B)_{(0,0)}}$.
Due to translational invariance, the only relevant part of the summation is the relative positioning of $B$ and $B'^\dagger$, thus reducing the computation to a single summation.
In the remaining part of this section, we define each environment relative to the position of the $B'^\dagger$ tensor in the bra layer, so that the computation of quantities can be completed by placing a pair of an $A$ and a $B'^\dagger$ tensor in the center.

The idea of the procedure is to compute new boundary tensors that approximate the infinite summations in the different regions of the environment.
For example, all terms that contain a $B$ tensor in the ket layer that is located strictly on the left side of the unit cell, can then be computed by summing over the following contractions:
\begin{equation}
  \begin{tikzpicture}[scale=0.8, transform shape]

    \begin{scope}[local bounding box=scope1, shift={(0,0)}]

      \begin{scope}[shift={(0,-2)}]
        \draw[doubline] (0,0.5) -- (0,-2.5);
        \draw[doubline] (1,0.5) -- (1,-2.5);
        \draw[doubline] (2,0.5) -- (2,-2.5);
        \draw[doubline] (3,0.5) -- (3,-2.5);

        \draw[doubline] (-0.5,0) -- (3.5,0);
        \draw[doubline] (-0.5,-1) -- (3.5,-1);
        \draw[doubline] (-0.5,-2) -- (3.5,-2);

        \node[tensor] at (0,0) {};
        \node[tensor] at (0,-1) {};
        \node[tensor] at (0,-2) {};
        \node[tensor] at (1,0) {};
        \node[tensorb] at (1,-1) {};
        \node[tensor] at (1,-2) {};
        \node[tensor] at (2,0) {};
        \node[tensor, draw=white] at (2,-1) {};
        \node[tensor] at (2,-2) {};
        \node[tensor] at (3,0) {};
        \node[tensor] at (3,-1) {};
        \node[tensor] at (3,-2) {};

        \node[] at (3.75,-1) {$\dots$};
        \node[] at (-0.75,-1) {$\dots$};
        \node[] at (2,0.8) {$\vdots$};
        \node[] at (2,-2.6) {$\vdots$};

        \node[] at (4.25,-1) {$+$};
      \end{scope}

      \begin{scope}[shift={(5.5,-2)}]
        \draw[doubline] (0,0.5) -- (0,-2.5);
        \draw[doubline] (1,0.5) -- (1,-2.5);
        \draw[doubline] (2,0.5) -- (2,-2.5);
        \draw[doubline] (3,0.5) -- (3,-2.5);

        \draw[doubline] (-0.5,0) -- (3.5,0);
        \draw[doubline] (-0.5,-1) -- (3.5,-1);
        \draw[doubline] (-0.5,-2) -- (3.5,-2);

        \node[tensor] at (0,0) {};
        \node[tensorb] at (0,-1) {};
        \node[tensor] at (0,-2) {};
        \node[tensor] at (1,0) {};
        \node[tensor] at (1,-1) {};
        \node[tensor] at (1,-2) {};
        \node[tensor] at (2,0) {};
        \node[tensor, draw=white] at (2,-1) {};
        \node[tensor] at (2,-2) {};
        \node[tensor] at (3,0) {};
        \node[tensor] at (3,-1) {};
        \node[tensor] at (3,-2) {};

        \node[] at (3.75,-1) {$\dots$};
        \node[] at (-0.75,-1) {$\dots$};
        \node[] at (2,0.8) {$\vdots$};
        \node[] at (2,-2.6) {$\vdots$};

      \end{scope}

      \begin{scope}[shift={(3,-6)}]
        \draw[line width=0.1cm] (0,0) -- (0,-2);
        \draw[line width=0.1cm] (0,0) -- (2,0);
        \draw[line width=0.1cm] (0,-2) -- (2,-2);
        \draw[line width=0.1cm] (2,0) -- (2,-2);
        \draw[doubline] (1,0) -- (1,-2);
        \draw[doubline] (0,-1) -- (2,-1);

        \node[N] at (0,0) {};
        \node[N] at (2,0) {};
        \node[N] at (2,-2) {};
        \node[N] at (0,-2) {};
        \node[N, T, rounded corners=0.15cm] at (1,0) {};
        \node[N, Th, rounded corners=0.15cm] at (2,-1) {};
        \node[N, T, rounded corners=0.15cm] at (1,-2) {};
        \node[B, Th, rounded corners=0.15cm] at (0,-1) {{\small $BT_4$ }};
        \node[tensor, draw=white] at (1,-1) {};

        \node[] at (-1,-1) {$\approx$};
        \node[] at (-2,-1) {$+~\cdots$};
      \end{scope}
    \end{scope}

  \end{tikzpicture}
\end{equation}
where we introduce the \emph{excitation environment} left row transfer matrix $B T_4$, represented by the colored shape.

These tensors can be computed in a very similar way to the standard norm environment boundary tensors by absorbing sites in an iterative way.
However, now the tensors contain multiple terms and all possible $B$ tensor locations have to be included.
The left row transfer matrix is updated by adding a pair of ground-state $A$ tensors to the $B T_4$ tensor of the previous iteration and adding the result to the contraction of the regular $T_4$ with a pair of a $B$ and an $A^\dagger$ tensor:
\begin{equation}
  \begin{tikzpicture}[scale=1, transform shape]

    \begin{scope}[local bounding box=scope1, shift={(0,0)}]
      \node[anchor=west] at (-0.5,1) {$BT_4'$};

      \begin{scope}
        \draw[line width=0.1cm] (0,0) -- (0,0.5);
        \draw[doubline] (0,0) -- (0.75,0);
        \draw[line width=0.1cm] (0,0) -- (0,-0.5);
        \node[B, Th] at (0,0) {};
        \node[] at (1.25,0) {$=$};
        \node[] at (1.55,0) {$~\Bigg($};
      \end{scope}

      \begin{scope}[shift={(2.25,0)}]

        \draw[line width=0.1cm] (0,0) -- (0,0.4);
        \draw[doubline] (0,0) -- (1,0);
        \draw[line width=0.1cm] (0,0) -- (0,-0.4);

        \draw[doubline] (1,0) -- (1,0.4);
        \draw[doubline] (1,0) -- (1.5,0);
        \draw[doubline] (1,0) -- (1,-0.4);

        \node[B, Th] at (0,0) {};
        \node[tensor] at (1,0) {};
        \node[] at (2,0) {$+$};

        \pic{projector};
        \pic{projector up};
      \end{scope}

      \begin{scope}[shift={(5.25,0)}]

        \draw[line width=0.1cm] (0,0) -- (0,0.4);
        \draw[doubline] (0,0) -- (1,0);
        \draw[line width=0.1cm] (0,0) -- (0,-0.4);

        \draw[doubline] (1,0) -- (1,0.4);
        \draw[doubline] (1,0) -- (1.5,0);
        \draw[doubline] (1,0) -- (1,-0.4);

        \node[N, Th] at (0,0) {};
        \node[tensorb] at (1,0) {};
        \node[] at (1.75,0) {$~\Bigg)$};
        \node[] at (2.35,0) {$\cdot e^{-i k_x}$};

        \pic{projector};
        \pic{projector up};
      \end{scope}

    \end{scope}

  \end{tikzpicture}
\end{equation}
Note that when adding terms with $B$ tensors located on different positions, their phase factors coming from the momentum superposition have to be taken into account.
For the resulting boundary tensors, only the relative phase between the terms they contain are important, and we can shift the overall phase by multiplying the boundary tensor by a factor $e^{i \phi}$.
In our implementation, we shift the overall phases of the boundary tensors after each iteration such that the phase of center site is always zero (corresponding to location $(0,0)$).

For the computation of the energy overlap matrix of~\Cref{eq:energy_matrix}, a second infinite summation must be performed; this time over all possible locations of the Hamiltonian.
Such summations come with more diagrams due to the support of the Hamiltonian on multiple sites~\footnote{We restrict ourselves in this work to nearest-neighbor Hamiltonians.
However, the Hamiltonian summation CTM scheme has previously been extended to next-nearest neighbor Hamiltonians and first applied in~\cite{niesenGroundstateStudySpin12018}.}, which have been worked out in the variational ground-state optimization algorithm~\cite{corbozVariationalOptimizationInfinite2016}.

Another type of boundary tensor now comes into play (denoted by the red shading) which contains summations over all possible Hamiltonian terms in the different regions.
All terms in the environment with combinations of a $B$ tensor and a Hamiltonian that are \emph{not} located in the same region and not on the center site can then be written as
\begin{equation}
  \begin{tikzpicture}[scale=1, transform shape]

    \begin{scope}[local bounding box=scope1, shift={(0,0)}]
      \begin{scope}[shift={(0,0)}]
        \draw[line width=0.1cm] (0,0) -- (0,-2);
        \draw[line width=0.1cm] (0,0) -- (2,0);
        \draw[line width=0.1cm] (0,-2) -- (2,-2);
        \draw[line width=0.1cm] (2,0) -- (2,-2);
        \draw[doubline] (1,0) -- (1,-0.8);
        \draw[doubline] (1,-2) -- (1,-1.2);
        \draw[doubline] (0,-1) -- (0.8,-1);
        \draw[doubline] (2,-1) -- (1.2,-1);

        \node[H] at (0,0) {};
        \node[N] at (2,0) {};
        \node[N] at (2,-2) {};
        \node[N] at (0,-2) {};
        \node[B, T] at (1,0) {};
        \node[N, Th] at (2,-1) {};
        \node[N, T] at (1,-2) {};
        \node[N, Th] at (0,-1) {};

        \node[] at (3,-1) {$+$};
      \end{scope}

      \begin{scope}[shift={(4,0)}]
        \draw[line width=0.1cm] (0,0) -- (0,-2);
        \draw[line width=0.1cm] (0,0) -- (2,0);
        \draw[line width=0.1cm] (0,-2) -- (2,-2);
        \draw[line width=0.1cm] (2,0) -- (2,-2);
        \draw[doubline] (1,0) -- (1,-0.8);
        \draw[doubline] (1,-2) -- (1,-1.2);
        \draw[doubline] (0,-1) -- (0.8,-1);
        \draw[doubline] (2,-1) -- (1.2,-1);

        \node[H] at (0,0) {};
        \node[B] at (2,0) {};
        \node[N] at (2,-2) {};
        \node[N] at (0,-2) {};
        \node[N, T] at (1,0) {};
        \node[N, Th] at (2,-1) {};
        \node[N, T] at (1,-2) {};
        \node[N, Th] at (0,-1) {};

        \node[] at (3,-1) {$+$};
        \node[] at (3.5,-1) {$\dots$};
      \end{scope}

    \end{scope}
  \end{tikzpicture}
\end{equation}
where the first diagram corresponds to terms with a Hamiltonian in the upper left corner and a $B$ tensor in the upper central column and the second diagram contains terms where now the $B$ is located in the upper right corner.

The remaining terms are those where both a Hamiltonian and a $B$ tensor are located in the \emph{same} region, so that they should be contained in a single boundary tensor.
This type of boundary tensor, which we designate by both color and shading together, can be computed by using the same ideas as we used for the previous types.
The update of the left row transfer matrix $EBT_4$, as an example, can be represented by the following diagram:
\begin{equation}
  \begin{tikzpicture}[scale=1, transform shape]

    \begin{scope}
      \node[anchor=west] at (-0.5,1) {$EBT_4'$};

      \begin{scope}
        \draw[line width=0.1cm] (0,0) -- (0,0.5);
        \draw[doubline] (0,0) -- (0.75,0);
        \draw[line width=0.1cm] (0,0) -- (0,-0.5);
        \node[HB, Th] at (0,0) {};
        \node[] at (1.25,0) {$=$};
        \node[] at (1.55,0) {$~\Bigg($};
        \node[] at (1.25,-2.5) {$+$};
      \end{scope}

      \begin{scope}[shift={(2.25,0)}]
        \node[Hconnecth] at (0.5,0) {};

        \draw[line width=0.1cm] (0,0) -- (0,0.4);
        \draw[doubline] (0,0) -- (1,0);
        \draw[line width=0.1cm] (0,0) -- (0,-0.4);

        \draw[doubline] (1,0) -- (1,0.4);
        \draw[doubline] (1,0) -- (1.5,0);
        \draw[doubline] (1,0) -- (1,-0.4);

        \node[B, Th] at (0,0) {};
        \node[tensorxin] at (0.15,0) {$\times$};
        \node[tensorx] at (1,0) {$\times$};
        \node[] at (2,0) {$+$};

        \pic{projector};
        \pic{projector up};
      \end{scope}

      \begin{scope}[shift={(5.25,0)}]
        \node[Hconnecth] at (0.5,0) {};

        \draw[line width=0.1cm] (0,0) -- (0,0.4);
        \draw[doubline] (0,0) -- (1,0);
        \draw[line width=0.1cm] (0,0) -- (0,-0.4);

        \draw[doubline] (1,0) -- (1,0.4);
        \draw[doubline] (1,0) -- (1.5,0);
        \draw[doubline] (1,0) -- (1,-0.4);

        \node[N, Th] at (0,0) {};
        \node[tensorxin] at (0.15,0) {$\times$};
        \node[tensorxb] at (1,0) {$\times$};

        \pic{projector};
        \pic{projector up};
      \end{scope}

      \begin{scope}[shift={(2.25,-2.5)}]
        \draw[line width=0.1cm] (0,0) -- (0,0.4);
        \draw[doubline] (0,0) -- (1,0);
        \draw[line width=0.1cm] (0,0) -- (0,-0.4);

        \draw[doubline] (1,0) -- (1,0.4);
        \draw[doubline] (1,0) -- (1.5,0);
        \draw[doubline] (1,0) -- (1,-0.4);

        \node[HB, Th] at (0,0) {};
        \node[tensor] at (1,0) {};
        \node[] at (2,0) {$+$};

        \pic{projector};
        \pic{projector up};
      \end{scope}

      \begin{scope}[shift={(5.25,-2.5)}]
        \draw[line width=0.1cm] (0,0) -- (0,0.4);
        \draw[doubline] (0,0) -- (1,0);
        \draw[line width=0.1cm] (0,0) -- (0,-0.4);

        \draw[doubline] (1,0) -- (1,0.4);
        \draw[doubline] (1,0) -- (1.5,0);
        \draw[doubline] (1,0) -- (1,-0.4);

        \node[H, Th] at (0,0) {};
        \node[tensorb] at (1,0) {};

        \node[] at (1.75,0) {$~\Bigg)$};
        \node[] at (2.35,0) {$\cdot e^{-i k_x}$};
        \pic{projector};
        \pic{projector up};
      \end{scope}

    \end{scope}  \end{tikzpicture}
\end{equation}

The crosses depict sites on which the physical indices are left open so that a Hamiltonian (red bar in the first line of diagrams) can be placed there.
This requires a special $T_4o$-type tensor that is equivalent to one that is used in the variational optimization algorithm for ground states~\cite{corbozVariationalOptimizationInfinite2016}.
The computation of these tensors is shown in~\Cref{sec:contraction_scheme}.

Lastly, there are terms in which the Hamiltonian is situated partially on the center site, for which the same $T_4o$-type tensors can be used: either the $EBT_4o$ tensor, if the $B$ tensor is located in the same sector as the Hamiltonian, or the $ET_4o$ tensor (containing only a Hamiltonian term), if the $B$ tensor is located elsewhere.
The $ET_4o$ tensor appears in the following terms:
\begin{equation}
  \begin{tikzpicture}[scale=1, transform shape]
    \begin{scope}[local bounding box=scope1, shift={(0,0)}]
      \begin{scope}[shift={(0,0)}]
        \node[Hconnecth] at (0.5,-1) {};

        \draw[line width=0.1cm] (0,0) -- (0,-2);
        \draw[line width=0.1cm] (0,0) -- (2,0);
        \draw[line width=0.1cm] (0,-2) -- (2,-2);
        \draw[line width=0.1cm] (2,0) -- (2,-2);
        \draw[doubline] (1,0) -- (1,-0.8);
        \draw[doubline] (1,-2) -- (1,-1.2);
        \draw[doubline] (0,-1) -- (0.8,-1);
        \draw[doubline] (2,-1) -- (1.2,-1);

        \node[B] at (0,0) {};
        \node[N] at (2,0) {};
        \node[N] at (2,-2) {};
        \node[N] at (0,-2) {};
        \node[N, T] at (1,0) {};
        \node[N, Th] at (2,-1) {};
        \node[N, T] at (1,-2) {};
        \node[N, Th] at (0,-1) {};
        \node[tensorxin] at (0.15,-1) {$\times$};
        \node[tensorx, draw=white] at (1,-1) {$\times$};

        \node[] at (3,-1) {$+$};
      \end{scope}

      \begin{scope}[shift={(4,0)}]
        \node[Hconnecth] at (0.5,-1) {};

        \draw[line width=0.1cm] (0,0) -- (0,-2);
        \draw[line width=0.1cm] (0,0) -- (2,0);
        \draw[line width=0.1cm] (0,-2) -- (2,-2);
        \draw[line width=0.1cm] (2,0) -- (2,-2);
        \draw[doubline] (1,0) -- (1,-0.8);
        \draw[doubline] (1,-2) -- (1,-1.2);
        \draw[doubline] (0,-1) -- (0.8,-1);
        \draw[doubline] (2,-1) -- (1.2,-1);

        \node[N] at (0,0) {};
        \node[N] at (2,0) {};
        \node[N] at (2,-2) {};
        \node[N] at (0,-2) {};
        \node[B, T] at (1,0) {};
        \node[N, Th] at (2,-1) {};
        \node[N, T] at (1,-2) {};
        \node[N, Th] at (0,-1) {};
        \node[tensorxin] at (0.15,-1) {$\times$};
        \node[tensorx, draw=white] at (1,-1) {$\times$};

        \node[] at (3,-1) {$+$};
        \node[] at (3.5,-1) {$\dots$};
      \end{scope}

    \end{scope}

  \end{tikzpicture}
\end{equation}

In~\Cref{sec:contraction_scheme} we show all relevant diagrams for the computation of the combined Hamiltonian and $B$ tensor boundary tensors~\footnote{Note that all diagrams are combinations of those used in the variational optimization algorithm for ground states~\cite{corbozVariationalOptimizationInfinite2016} and can be easily adapted from an existing code.}.
Once all boundary tensors have been computed, all terms in the evaluation of the energy and norm overlap matrices can be calculated by using the appropriate boundary tensors for all possible placements of Hamiltonian and $B$ tensors. 

While the computational cost of the individual CTM iterations scales similarly to the cost of the variational ground state algorithm, the basis size of the overlap matrices also increases with $D$.
In practice, we make sure in our simulations to perform enough CTM iterations to achieve convergence.
Generally, the convergence of the CTM for the excitations is similar to the convergence of the CTM for the ground state that we start from.
For large scale simulations, the eigenvalue problem of~\Cref{eq:main_eig_problem} can be solved with an implicit iterative solver, since generally only the few lowest eigenvalues are relevant.
 
\subsubsection{Arbitary unit cell sizes}
\label{ssub:general_unit_cell_size}

In the framework of CTM it is straightforward to extend the contractions to unit cells that are larger than a single site~\cite{corbozStripesTwodimensionalModel2011,corbozCompetingStatesModel2014}, for models that partially break translational invariance.
Keeping track of separate environments for each site in the unit cell, the computation of an expectation value consists of separate contractions of each site tensor with its respective environment.

Again the exited states, now parameterized by a vector $\va{B}$ containing the elements of all tensors $B[1] \dots B[n]$, with $n$ the number of sites within the unit cell, are restricted to those that are orthogonal to the ground state.
If $X_{B[i]}=\mathrm{null}\qty(\bra{\Psi}\pdv{B[i]}\ket{\Phi\qty(B)_k})$ contains a basis of vectors $\va{B}^m[i]$ that forms the null space of the ground-state environment (reshaped to a $dD^4$ vector) of unit cell site $i$, then a complete basis for the excitation parameters $\va{B}$ is formed by $X_{B[1]} \bigoplus \dots \bigoplus X_{B[n]}$.
In this formulation it is clear that the number of free parameters describing the excited state scales linearly with the number of sites in the unit cell.

If the underlying ground state does not break the translational symmetry fully within the unit cell, for example a N\'eel pattern inside a $2 \times 2$ unit cell, the excitation ansatz for a given momentum $k$ can also represent reflections in the Brillouin zone at $k+\pi$ in either direction.
Restricting the tensors to a certain pattern excludes reflections from other momenta, while also reducing the computational cost of each CTM iteration and reducing the required number of basis tensors for the overlap matrices.

\subsubsection{Exploiting symmetries}
\label{ssub:symmetric_excitations_ctm}

Of great importance in many tensor network simulations is the ability to impose certain symmetries on the states, since it greatly reduces the number of free parameters in the ansatz and therefore improves numerical stability and speeds up calculations~\cite{bauerImplementingGlobalAbelian2011,singhTensorNetworkStates2011a}.
Additionally, it enables optimization algorithms to target a specific symmetry sector, which is useful for many physical ground states.
As has been shown in the one-dimensional case~\cite{zauner-stauberTopologicalNatureSpinons2018}, symmetries can be also used effectively in simulations of excited states, classified by their quantum number difference to the ground state.

In our implementation, we impose finite Abelian group symmetries $\mathbb{Z}_n$ as well as $U(1)$ symmetry, which we make use of in our simulations described in~\Cref{sec:benchmark_results}.
Any excited state that is constrained within a different symmetry sector than the ground state is automatically orthogonal to the ground state; therefore the choice of basis for the $B$ tensors in this case is arbitrary.

\subsubsection{Fermionic systems}
\label{ssub:fermionic_excitations_ctm}

Fermionic systems have also been studied using 2D tensor network methods, with a computational cost that is equivalent to bosonic systems~\cite{corbozSimulationInteractingFermions2010,krausFermionicProjectedEntangled2010,pinedaUnitaryCircuitsStrongly2010a,barthelContractionFermionicOperator2009,shiGradedProjectedEntangledPair2009,corbozFermionicMultiscaleEntanglement2009,pizornFermionicImplementationProjected2010,guGrassmannTensorNetwork2010,corbozSimulationStronglyCorrelated2010}.
Here we use the same ideas to extend our method to simulate quasiparticle excitations in fermionic systems.

While we refer to Refs.~\cite{corbozFermionicMultiscaleEntanglement2009,corbozSimulationStronglyCorrelated2010} for details, the implementation involves imposing $\mathbb{Z}_2$ symmetry to preserve fermionic parity and the introduction of \emph{swap} tensors whenever two lines cross in the two-dimensional projection of the tensor contractions, to account for the fermionic anticommutation rules.
A quasiparticle excitation can be represented by $B$ tensors of either even parity, consisting of an even number of fermionic creation and annihilation operators, or odd parity, relative to the ground state.
Tensors of odd parity can be constructed by adding an index of dimension 1 that carries an odd quantum number to an even parity tensor.

\section{Results}
\label{sec:benchmark_results}

\subsection{Transverse field Ising model}
\label{sub:transverse_ising}

\begin{figure}[!t]
  \includegraphics[width=\linewidth]{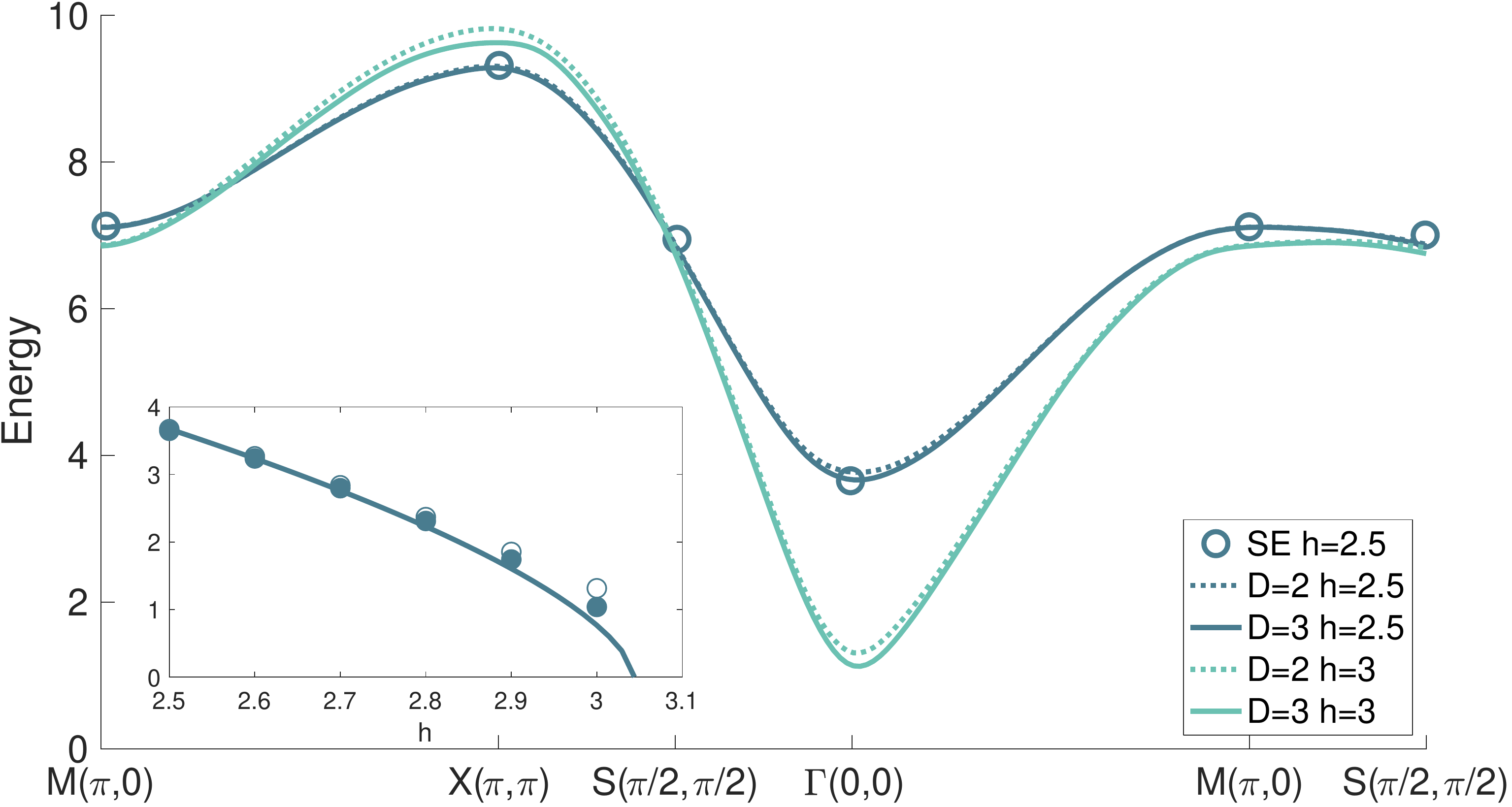}
  \caption{Dispersion of the transverse field Ising model with iPEPS (D=2,3) for field strength $h=2.5$ and $h=3$. The circles are series expansion results~\cite{oitmaa2006series}. The inset shows a comparison of the gap as a function of $h$ between the $D=2$ (open circles) and $D=3$ (filled circles) iPEPS results and the analytical scaling relation $\Delta E \propto \abs{h-h_{crit}}^{\nu}$, showing improvement with increasing bond dimension.}
  \label{fig:ising_var_h_lowest_exci}
\end{figure}

The 2D quantum ferromagnetic Ising model with transverse magnetic field of strength $h$ is described by the Hamiltonian

\begin{equation}
  \mathrm{H} = -J\sum_{<i,j>} \sigma^x_i \sigma^x_j - h \sum_i \sigma^z_i
\end{equation}
with $J=1$.
The ground state of this model can be accurately simulated with iPEPS throughout the phase diagram~\cite{vanderstraetenGradientMethodsVariational2016}, which contains a symmetry-broken phase and a polarized phase with a transition at $h_{crit} = 3.04438(2)$~\cite{bloteClusterMonteCarlo2002}.

We use the iPEPS excitations method to compute the dispersion of the lowest-lying excitation, a magnon, for $h=2.5,3$ on a path through high-symmetry points of the Brillouin zone, as shown in~\Cref{fig:ising_var_h_lowest_exci}.
The results for $h=2.5$ show already convergence in the bond dimension for $D=2,3$ and agree well with values from series expansions (SE)~\cite{oitmaa2006series}.
For $h=3$, being closer to the critical point, we observe stronger dependence on the bond dimension around the $X(\pi,\pi)$ and $\Gamma(0,0)$ points, but we do find a systematic improvement with increasing bond dimension, as show in the inset of~\Cref{fig:ising_var_h_lowest_exci} for the $\Gamma$ point.
Our results correspond well to those of earlier iPEPS calculations~\cite{vanderstraetenSimulatingExcitationSpectra2019}, demonstrating that our CTM-based contraction method performs equivalently to the method used in that work.

\begin{figure}[!t]
  \includegraphics[width=\linewidth]{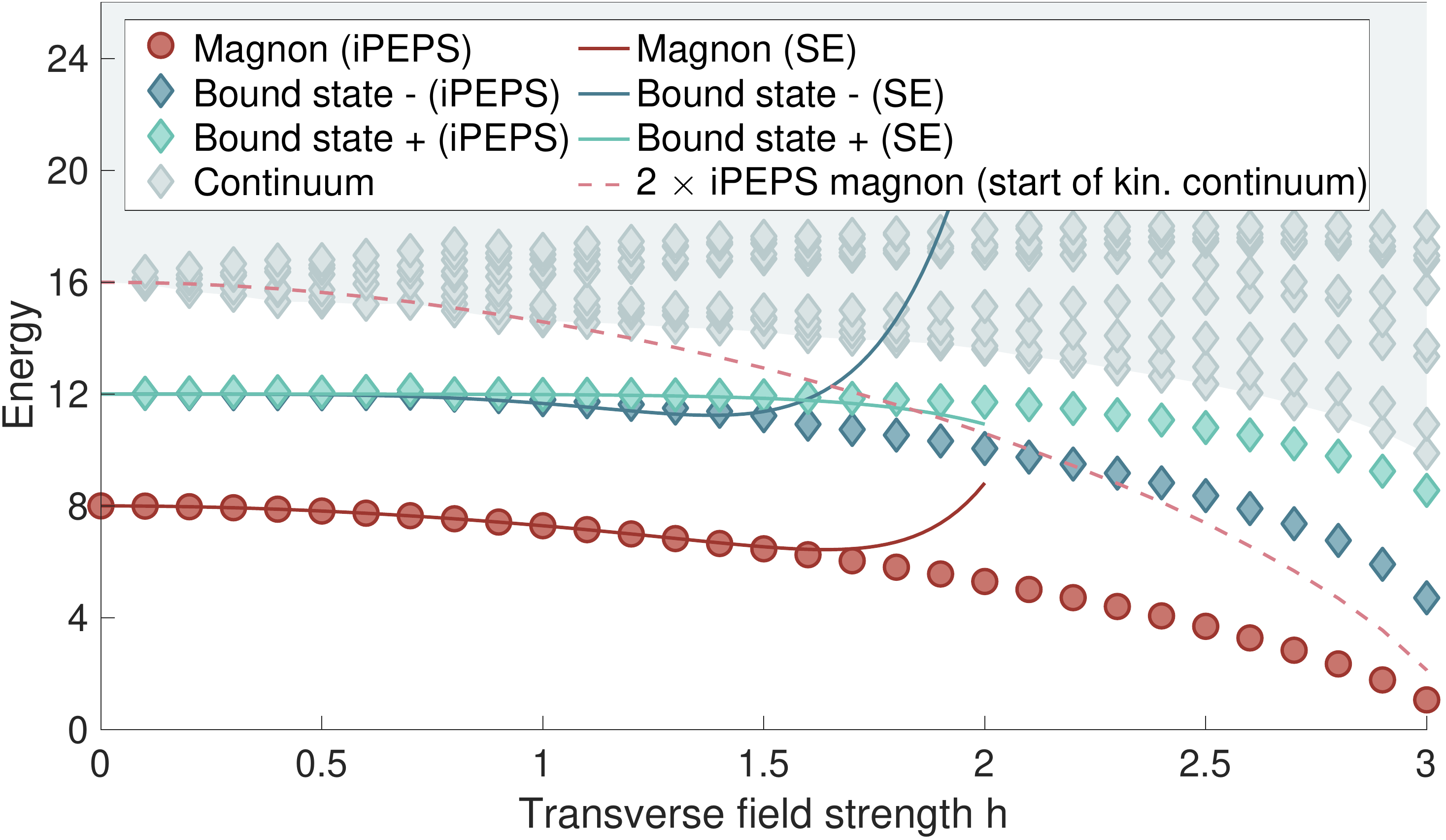}
  \caption{iPEPS ($D=3$) results for the three lowest-lying excited states as a function of field strength $h$, compared to series expansion results. The red dashed line represents the start of the kinematic continuum.}
  \label{fig:ising_bound_states}
\end{figure}

Another interesting aspect of the excitations in the Ising model is the appearance of bound states of two magnons below the continuum.
For $h=0$, these states are neighboring pairs of spin flips with energy $12J$ instead of the $16J$ energy of two non-interacting free magnons, and we can trace their energies for $h \to h_{crit}$.
In \Cref{fig:ising_bound_states}, the energies of the lowest excitation (magnon) mode and the two bound modes ($-,+$) are plotted.
We can compare our results in~\Cref{fig:ising_bound_states} to series expansion results~\cite{dusuelBoundStatesTwodimensional2010} and we observe close agreement in the region of small $h$, where the series expansions are accurate.

While in the $h=0$ case the Hamiltonian contains no terms that couple the different two-magnon bound states, leaving them completely degenerate at energy $12J$, the two energies split for $h>0$.
This energy difference is clearly visible in our results, showing that the iPEPS representation is able to account for such effects.
We also show several higher-lying eigenvalues which correspond to multi-particle states within a continuum.
Although the ansatz is by construction only suitable for describing single-particle states, the eigenvalues in the continuum become increasingly spread with larger $h$, showing level repulsion effects within the continuum. 

\subsection{Heisenberg model}
\label{sub:heisenberg}

We now focus on another model to demonstrate our framework: the 2D spin-$\sfrac{1}{2}$ quantum antiferromagnetic Heisenberg model, defined as
\begin{equation}
  \mathrm{H} = J \sum_{<i,j>} S^z_i S^z_{j} + \lambda \qty(S^x_i S^x_{j} + S^y_i S^y_{j})
  \label{eq:heis_ham}
\end{equation}
with $J=1$ and $\lambda=1$.

For our ground state, we enforce the U(1) symmetry that corresponds to conservation of the total $z$-component of the spin, $S^z_{tot}$, and we use a $2 \times 2$ unit cell with a checkerboard pattern.
Since the ground state sponteously breaks the SU(2) symmetry of the Hamiltonian, the system exhibits gapless excitations.
The ground-state tensors are fixed in the $S^z_{tot} =0$ sector and in the following sections we consider excitations in the $S^z_{tot}=1$ sector, corresponding primarily to a magnon mode.

\subsubsection{Dispersion}
\label{ssub:dispersion}

\begin{figure}[!t]
  \includegraphics[width=\linewidth]{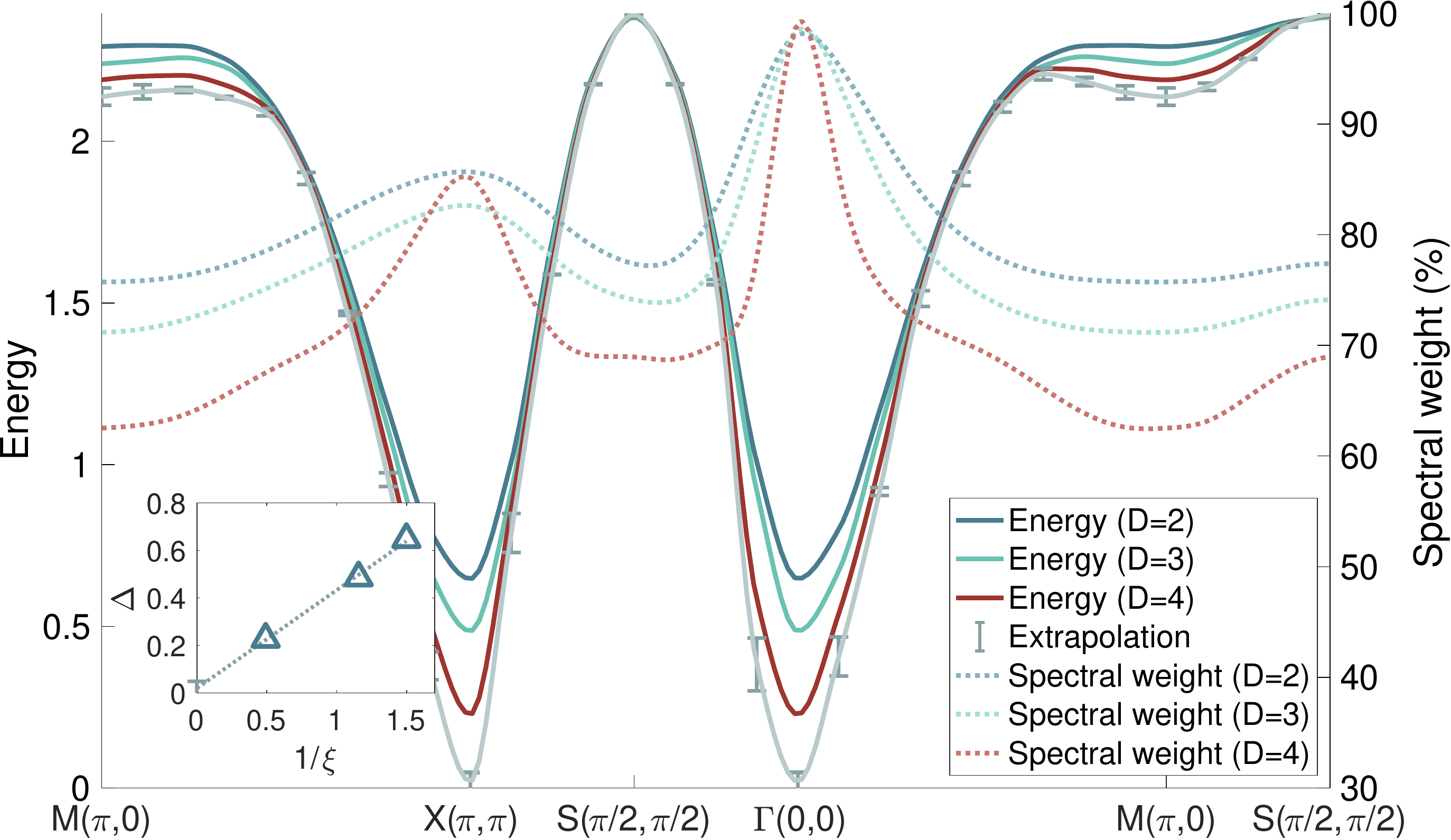}
  \caption{Results for the dispersion of the Heisenberg model along a representative path through the Brillouin zone for different bond dimension values, with extrapolations in terms of the inverse correlation length. The dotted lines show the relative spectral weight of the excitations. In the inset the value of the artificial gap $\Delta$ at $k=(\pi,\pi)$ is plotted as a function of the inverse correlation length.}
  \label{fig:heis_bril_exci}
\end{figure}

We plot the dispersion of the magnon excitation for several values of the bond dimension in~\Cref{fig:heis_bril_exci}.
The most obvious dependence on $D$ is around the gapless points $X(\pi,\pi)$ and $\Gamma(0,0)$, and around $M(\pi,0)$, which we discuss in the remainder of this section.
In other regions of the Brillouin zone the energies are already well converged in $D$.
Observe that there is a finite energy at the gapless points which decreases with increasing~$D$.
This is consistent with the findings in Ref.~\cite{corbozFiniteCorrelationLength2018} that the finite $D$ effectively introduces a finite correlation length in the ground state of the 2D Heisenberg model, which will only diverge in the infinite-$D$ limit.
This effective correlation length can be used for accurate extrapolations by using a scaling ansatz reminiscent of the finite size scaling often used in numerical simulations, an idea that has been applied in the context of MPS~\cite{tagliacozzoScalingEntanglementSupport2008,pollmannTheoryFiniteEntanglementScaling2009,pirvuMatrixProductStates2012}, classical 2D systems~\cite{nishinoNumericalRenormalizationGroup1996} and recently also 2D iPEPS~\cite{corbozFiniteCorrelationLength2018,raderFiniteCorrelationLength2018}.
In the inset of~\Cref{fig:heis_bril_exci} we show the dependence of the artificial gap at $k=(\pi,\pi)$ of the lowest excited state on the inverse effective correlation length of the ground state and find that a linear extrapolation yields a value compatible with a vanishing gap, suggesting that also for excitations the correlation length is a useful quantity for extrapolations.

The dispersion on the line between $(\pi,0)$ and $(\pi/2,\pi/2)$ has been the topic of much research, since the conventional linear spin wave theory - predicting a flat dispersion - is contradicted by numerical results as well as experiments on quantum antiferromagnets.
Regarding the nature of the excitations around $(\pi,0)$, which we discuss in~\Cref{ssub:nature_of_excitations}, several theories have been proposed, such as an interaction between the magnon and a double spinon mode~\cite{dallapiazzaFractionalExcitationsSquarelattice2015,shaoNearlyDeconfinedSpinon2017a,christensenQuantumDynamicsEntanglement2007} or a repulsion from continua of multi-magnon (bound) states~\cite{verresenQuantumDynamicsSquarelattice2018}.
We indeed observe a dip in the magnon energy, increasing with bond dimension, in agreement with earlier iPEPS results~\cite{vanderstraetenSimulatingExcitationSpectra2019}.

In~\Cref{fig:heis_swa_exci}, we zoom in to the region between $M(\pi,0)$ and $S(\pi/2,\pi/2)$ and compare our results to other numerical, analytical and experimental results~\cite{dallapiazzaFractionalExcitationsSquarelattice2015,christensenQuantumDynamicsEntanglement2007,zhengSeriesStudiesSpin2005,sandvikHighEnergyMagnonDispersion2001}.
Clearly the dependence on the bond dimension is stronger at $M$ than at $S$, with the $D=4$ result approaching the series expansion and quantum Monte Carlo results.
Although our results are in close agreement with iPEPS results in Ref.~\cite{vanderstraetenSimulatingExcitationSpectra2019} for equal bond dimensions, small deviations could be attributed to the fact that here we impose the $U(1)$ symmetry, which restricts the excitations to a fixed sector.

\subsubsection{Dynamical structure factor}
\label{ssub:dynamical_structure_factor}

\begin{figure}[!t]
  \centering
  \includegraphics[width=\linewidth]{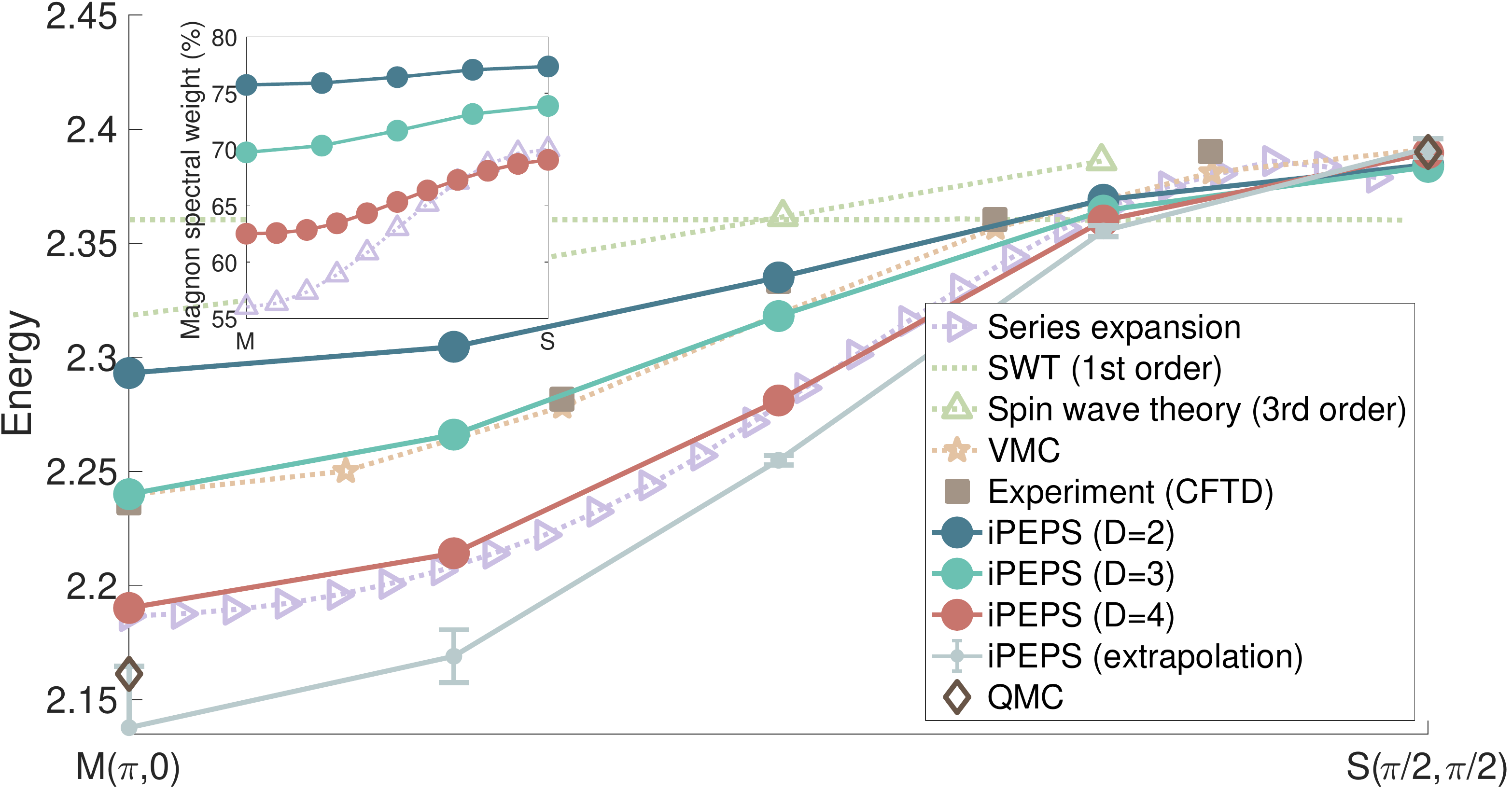}
  \caption{Heisenberg dispersion between M$(\pi,0)$ and S$(\pi/2,\pi/2)$. iPEPS results (D=3,4) are shown along with results from various methods as well as experiments~\cite{dallapiazzaFractionalExcitationsSquarelattice2015,christensenQuantumDynamicsEntanglement2007,zhengSeriesStudiesSpin2005,sandvikHighEnergyMagnonDispersion2001}. The inset shows the spectral weight of the lowest excitation mode relative to the static structure factor, compared to series expansion results.}
  \label{fig:heis_swa_exci}
\end{figure}

An important quantity for the low-energy behavior of quantum systems is the dynamical structure factor, which is defined, for spin systems, in terms of dynamical correlation functions as
\begin{equation}
  \mathcal{S}^{\alpha \beta}(k,\omega) = \frac{1}{2 \pi} \int_{-\infty}^{\infty} dt~e^{i \omega t} \expval{S^{\alpha}_{-k}(t) S^{\beta}_{k}(0)}
\end{equation}
with $\alpha,\beta=x,y,z$.
For excitations in the $S^z_{tot}=+1$ sector, the relevant version is in the transverse channel
\begin{equation}
  \mathcal{S}^{\mathrm{trans}}(k,\omega) = \frac{1}{2 \pi} \int_{-\infty}^{\infty} dt~e^{i \omega t} \expval{S^+_{-k}(t) S^-_{k}(0)}~.
\end{equation}
The relative spectral weight of the lowest-lying excited state can be computed as
\begin{equation}
  w = \frac{\qty|\bra{\Phi(B)_k}S^+_{k}\ket{\Psi(A)}|^2}{\int \dd{\omega} \mathcal{S}(k,\omega)}
\end{equation}
and is plotted in~\Cref{fig:heis_bril_exci} (dotted lines) and in the inset of~\Cref{fig:heis_swa_exci}.
The quantity in the denominator $\int \dd{\omega} \mathcal{S}(k,\omega)=\ev{S^+_{-k}(t=0) S^-_k(t=0)}$, the static structure factor, can be computed accurately with our contraction method even though the individual multi-particle states cannot be represented.
At the location of the local minimum in the dispersion at $M$, the spectral weight of the magnon mode has been found to decrease significantly~\cite{shaoNearlyDeconfinedSpinon2017a,zhengSeriesStudiesSpin2005}, compared to the value at $S$ (from $70\% \to 40\%$).
Though we do observe a decrease in spectral weight, the difference is less pronounced ($70\% \to 63\%$ for $D=4$), which is likely an effect of the finite bond dimension, since we see clear improvement as $D$ increases.

\subsubsection{Nature of excitations at $(\pi,0)$}
\label{ssub:nature_of_excitations}

To further investigate the nature of the excitations at the $(\pi,0)$ point, we visualize the spin correlations within a single term in the momentum superposition.
At the center site in the figure, we exchange the ground-state pair of $\qty{A,A^\dagger}$ tensors for the optimized $\qty{B,B^\dagger}$ excitation tensors and compute local spin expectation values on other sites in its vicinity.
As expected for an excitation in the $S_z^{tot}=-1 (+1)$ sector, a site with up (down) magnetization in the ground state is flipped down (up) if a $B$ tensor is placed on that site, which affects neighboring sites due to entanglement.

\begin{figure}[!t]
  \includegraphics[width=\linewidth]{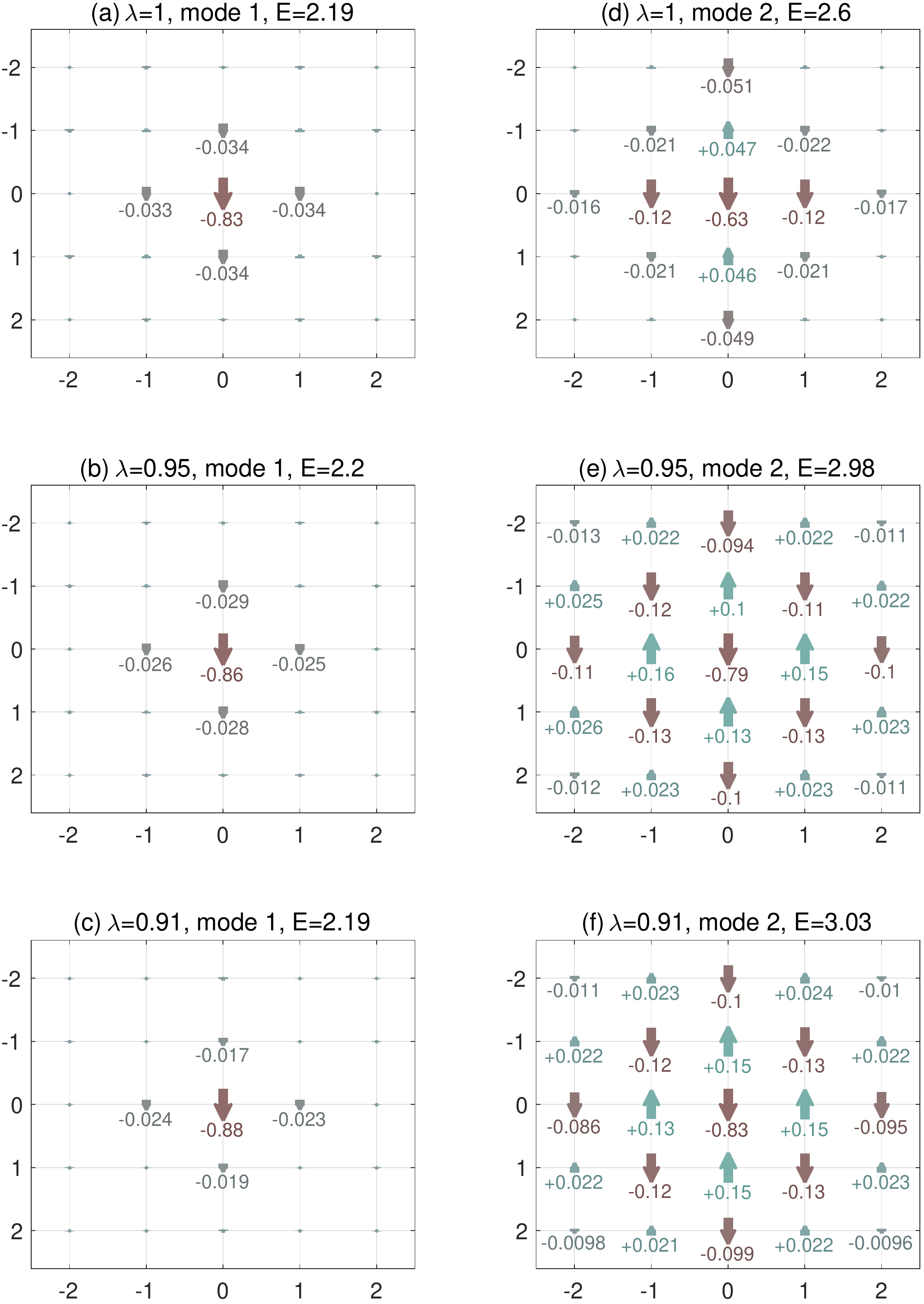}
  \caption{Real-space visualization of excited state at $k=(\pi,0)$, where a single $B$ tensor is placed on the center site and $\ev{S^z}$ is measured on a $5\times5$ patch. The arrows correspond to the difference in local magnetic moment with respect to the ground state, with the color representing a positive (negative) change. (a-c)~Lowest excited state (mode 1) at $\lambda=1,0.95,0.91$. (d)~Second-lowest excitation (mode 2) at the isotropic point. (e,f)~Second-lowest excitations away from the isotropic point, showing strong correlations on neighbouring sites that are part of three-magnon bound states that include the center site.}
  \label{fig:heis_exci_states_ani_comp_pi_0}
\end{figure}

A recent study~\cite{verresenQuantumDynamicsSquarelattice2018}, which used time evolution of semi-infinite cylindrical systems to obtain the dynamical structure factor, proposed a simple description of the excitation nature at the $M$ point; by moving away from the isotropic point of the Hamiltonian, i.e. $\lambda < 1$ in~\Cref{eq:heis_ham}, where the case $\lambda=0$ corresponds to the Ising limit.
It was observed that for small $\lambda$ three distinct types of resonances in the transverse structure factor could be identified: an isolated single magnon branch, three-magnon bound states and combinations of a magnon and a two-magnon bound state.
As $\lambda \to 1$, the multimagnon continuum moves smoothly towards the single magnon branch, until the magnon is no longer isolated at the isotropic point.

We observe that for $\lambda<1$ the real-space pictures of the lowest-lying excitations, shown in~\Cref{fig:heis_exci_states_ani_comp_pi_0}, vary continuously from the one at the isotropic point, without any qualitative difference.
From this point of view, the dip in energy around $M(\pi,0)$ could be understood as a result of the multimagnon continua moving close to the single magnon branch, which we observe in agreement with~\cite{verresenQuantumDynamicsSquarelattice2018}, as $\lambda \to 1$, pushing the magnon to a lower energy.
The effect of avoided crossing between a single magnon and a continuum has been observed in experiments~\cite{plumbQuasiparticlecontinuumLevelRepulsion2016} and numerical simulations~\cite{verresenAvoidedQuasiparticleDecay2019}, and another manifestation of the fact that repulsion between two bound modes can be captured using iPEPS has been shown in~\Cref{sub:transverse_ising}.

Slightly away from the isotropic point (\Cref{fig:heis_exci_states_ani_comp_pi_0}(d)~\footnote{Close to the isotropic point the close competition with the continuum of other states makes separating these bound states no longer possible.}), we can accurately identify the three-magnon bound states.
Although the bound states have equal energy in the noninteracting limit, level repulsion effects due to their coupling result in a superposition of such states having lower energy, which shows up as a single eigenvalue in our effective energy overlap matrix instead of several degenerate eigenvalues.
Such a state is consistent with the pictures of~\Cref{fig:heis_exci_states_ani_comp_pi_0}(e,f), and the energy at $\lambda=0.91$ is in agreement with Ref.~\cite{verresenQuantumDynamicsSquarelattice2018}.

While alternative explanations cannot be ruled out on these results alone, the ability of iPEPS to capture the nontrivial dip in the dispersion already from $D=2$, where any delocalized multiparticle states, for example deconfined spinons, would be especially hard to describe, suggests that the simple picture featuring a locally three-magnon bound state may provide a valid explanation.

\begin{figure}[!t]
  \includegraphics[width=\linewidth]{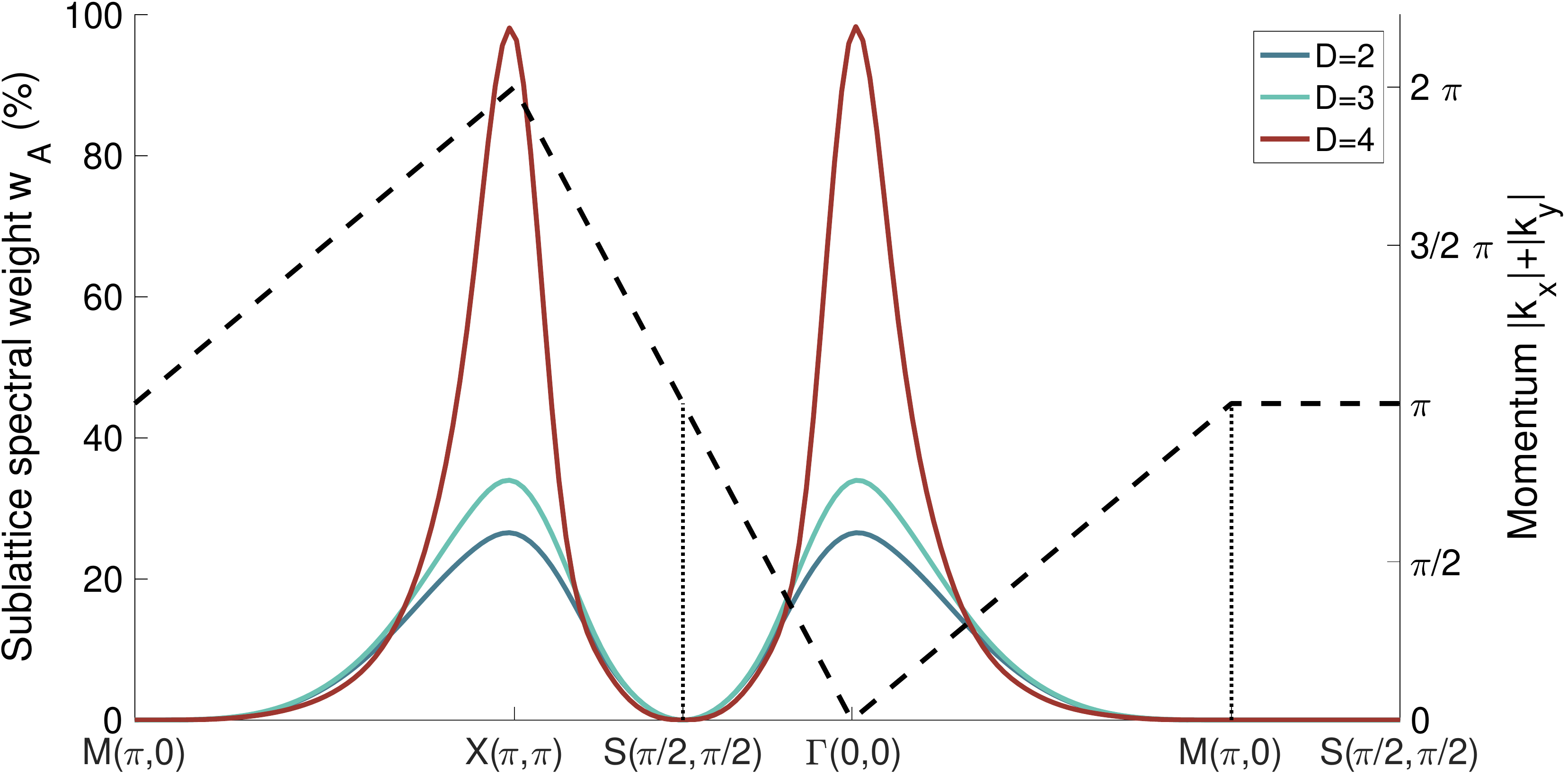}
  \caption{Sublattice spectral weight of the lowest excited state, acting with $S^-$ only on the spin-down sublattice (with respect to the ground state). The dashed line corresponds to the value of the sum of momenta. The absence of weight on the line $\abs{k_x} + \abs{k_y} = \pi$ implies that the magnon excitation is localized on one sublattice.} 
  \label{fig:heis_spec_w_A_to_B}
\end{figure}

In addition, it was observed in Ref.~\cite{verresenQuantumDynamicsSquarelattice2018} that the creation of magnons associated with one sublattice on the other sublattice (e.g. acting with a $S^-$ operator on the sublattice with negative ground-state magnetization, which affects mostly its neighboring sites), is suppressed on the line $\abs{k_x} + \abs{k_y} = \pi$ due to destructive interference.
This can be measured using a variant of the transverse spectral function, where the spin operator only acts on one sublattice:
\begin{equation}
  \mathcal{S}^{\mathrm{trans}}_{A}(k,\omega) = \sum_\alpha \delta(\omega - \omega_\alpha) ~ \qty|\bra{\alpha}S^-_{A,k}\ket{0}|^2
\end{equation}
and the corresponding sublattice spectral weight
\begin{equation}
  w_A = \frac{\qty|\bra{\Phi(B)_k}S^-_{A,k}\ket{\Psi(A)}|^2}{\int \dd{\omega} \mathcal{S}_{A}(k,\omega)}
\end{equation}
where $S^-_{A,k} = \sum_{x \in A} e^{i k x} S^-_x$ acts only on the $A$ (down) sublattice and $\alpha$ labels single-particle states.
In~\Cref{fig:heis_spec_w_A_to_B} we plot the spectral weight of the lowest excitation for $D=3,4$ and observe that it vanishes only on the line $\abs{k_x} + \abs{k_y} = \pi$, in qualitative agreement with finite-width cylinder simulations~\cite{verresenQuantumDynamicsSquarelattice2018}.

\subsection{Free spinless fermions}
\label{sub:free_spinless_fermions}

A powerful aspect of the iPEPS ansatz is that it is able to accurately capture ground states of fermionic systems, without the sign problem of quantum Monte Carlo.
We extend the excitations framework with the ability to treat fermionic systems, and we demonstrate its accuracy on a simple model of free fermions in the presence of a pairing term:
\begin{equation}
  \mathrm{H} = \sum_{<ij>} c^{\dagger}_i c_j + h.c - \gamma \left( c^{\dagger}_i c^{\dagger}_j + c_i c_j \right) - 2 \lambda \sum_i c^{\dagger}_i c_i~.
\end{equation}

In momentum space, the Hamiltonian takes the following form:
\begin{equation}
  \mathrm{H} = \sum_k -2 t_k ~ c^{\dagger}_k c_k + i \Delta_k \left( c^{\dagger}_k c^{\dagger}_{-k} - c_{-k} c_k \right)
\end{equation}
with $t_k \equiv \lambda - \mathrm{cos}~k_x - \mathrm{cos}~k_y$ and $\Delta_k \equiv \gamma \left( \mathrm{sin}~k_x + \mathrm{sin}~k_y \right)$.

This model can be exactly solved by a Bogoliubov transformation, which yields the lowest excited state
\begin{equation}
  \ket{\Phi_k} = d^{\dagger}_k \ket{\Psi_0},~ d^{\dagger}_k \equiv u_k c^{\dagger}_k + v_k c_{-k}
\end{equation}
of a single Bogoliubov mode on top of a vacuum state.
For $\abs{\lambda}>2$, the model is in a gapped phase, while for $\abs{\lambda} \leq 2$ the system is gapless.

\begin{figure}[!t]
  \includegraphics[width=\linewidth]{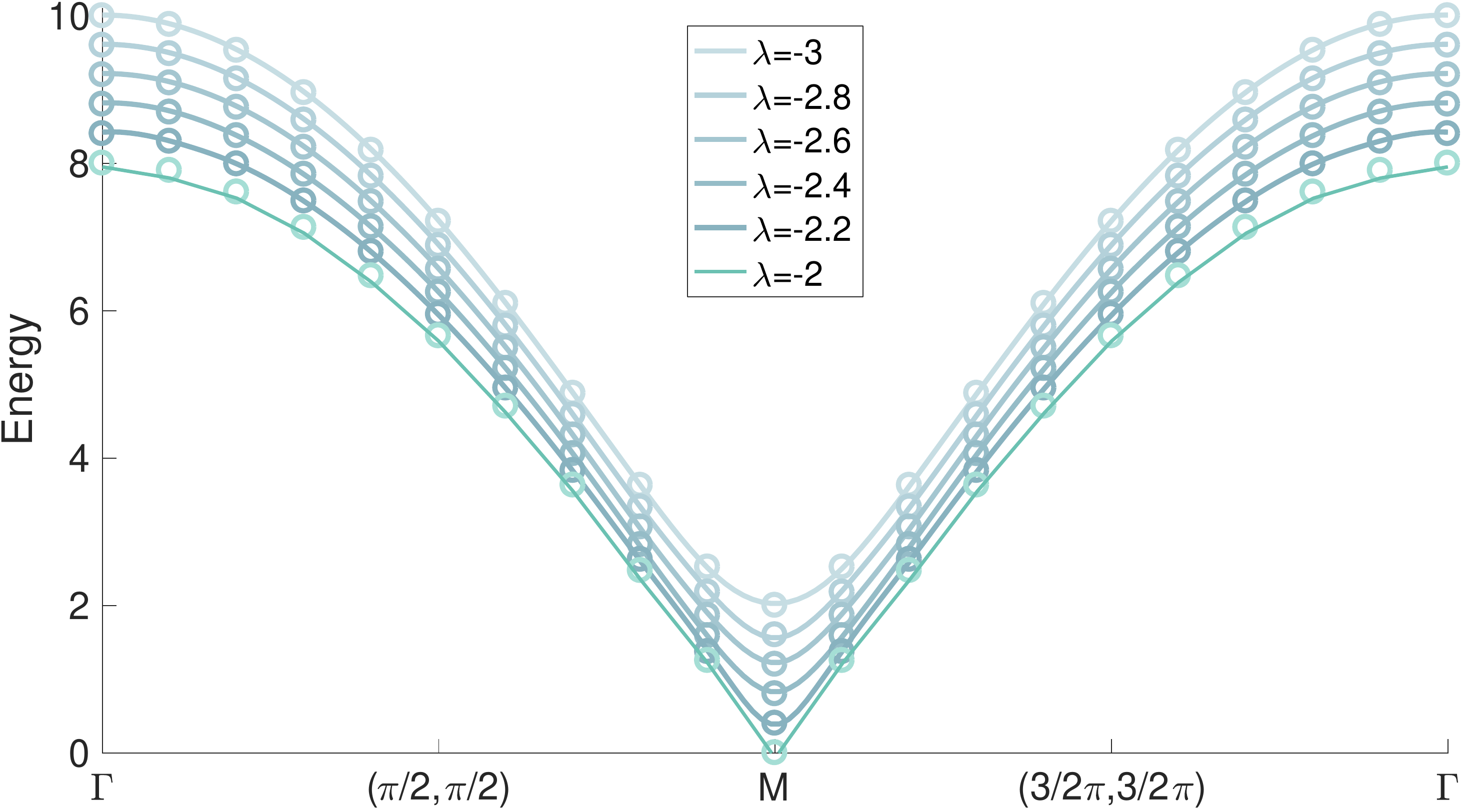}
  \caption{iPEPS results (solid lines) of the lowest excited state of a free fermionic model in a gapped phase. Circles represent exact values.}
  \label{fig:ff_D3_exci}
\end{figure}

In~\Cref{fig:ff_D3_exci}, we compare iPEPS results to the exact excitation energies, for several values of the chemical potential $\lambda$ with $\gamma=1$.
We observe excellent accuracy for $D=3$, especially around the minimum, also when approaching the gapless phase $\abs{\lambda} \leq 2$.
However, in the gapless phase the results are less accurate, partly because the ground state itself is more difficult to represent with iPEPS~\cite{corbozSimulationStronglyCorrelated2010} (but we do observe an improvement with $D$).
Indeed the $\lambda=-2$ results show more deviation from the exact values, and the energy at $M$ becomes slightly negative, which is a consequence of inaccuracy in the ground state that leads to an underestimation of the energy difference.

\section{Conclusions}
\label{sec:conclusions}

We have introduced an extension of the well established CTM method to simulate low-lying excited states with a single elementary excitation nature on top of a strongly correlated ground state.
This excitation ansatz has been used extensively in one-dimensional systems, in the context of matrix product states, and has previously been successfully extended to two-dimensional systems~\cite{vanderstraetenExcitationsTangentSpace2015,vanderstraetenSimulatingExcitationSpectra2019}, using a different approach to performing the necessary contractions than we use here.
We show that the CTM framework is equally capable of accurately computing dispersions in spin models and naturally allows for extensions to larger unit cell sizes.
Additionally, the implementation of symmetries reduces the computational cost and enables simulations that target excitations in specific symmetry sectors.
Lastly, a generalization to fermionic systems, based on earlier applications in tensor networks, is tested on a free fermionic system.
This leads the way to the study of more complex fermionic models, which would be of great importance in research areas such as high-T$_c$ superconductivity.

Our results on the transverse field Ising model and the Heisenberg model show close agreement with earlier iPEPS results~\cite{vanderstraetenSimulatingExcitationSpectra2019}.
We observe a finite gap due to a finite bond dimension at the gapless points of the Heisenberg model, however an extrapolation in the inverse effective correlation length suggests that the results are compatible with a vanishing gap in the infinite bond dimension limit.
Additionally, we find results at the $M$ point that are compatible with the spin wave anomaly, and real-space visualizations slightly away from the isotropic Heisenberg point show states that are compatible with three-magnon bound states that have been found in simulations on cylinders~\cite{verresenQuantumDynamicsSquarelattice2018}.

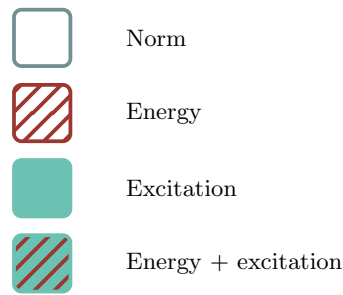
\begin{figure}[b!]
    \begin{tikzpicture}[scale=1, transform shape]
      \node[N] at (0,0) {};
      \node[anchor=west] at (1,0) {Norm};
      \node[H] at (0,-1) {};
      \node[anchor=west] at (1,-1) {Energy};
      \node[B] at (0,-2) {};
      \node[anchor=west] at (1,-2) {Excitation};
      \node[HB] at (0,-3) {};
      \node[anchor=west] at (1,-3) {Energy + excitation};
    \end{tikzpicture}
    \caption{The various types of environments and their coloring in this appendix.}
    \label{fig:diag_colors}
\end{figure}

\def\phases{1}

\begin{figure*}
  \begin{tikzpicture}[scale=0.84, transform shape]

    \begin{scope}[local bounding box=scope1, shift={(0,0)}]
      \node[anchor=west] at (-1.8,0) {$EBC_1'$};

      \begin{scope}
        \draw[line width=0.1cm] (0,0) -- (0,-.75);
        \draw[line width=0.1cm] (0,0) -- (0.75,0);
        \node[HB] at (0,0) {};
      \end{scope}

      \begin{scope}[shift={(1.25,0)}]
        \node[] at (0,0) {$=$};
      \end{scope}

      \begin{scope}[shift={(2.25,0)}]
        \draw[line width=0.1cm] (0,0) -- (0,-.4);
        \draw[line width=0.1cm] (0,0) -- (1,0);
        \draw[doubline] (1,0) -- (1,-0.4);
        \draw[line width=0.1cm] (1,0) -- (1.5,0);
        \node[HB] at (0,0) {};
        \node[N, T] at (1,0) {};
        \node[] at (2,0) {$+$};

        \pic{projector};
      \end{scope}

      \begin{scope}[shift={(5.25,0)}]
        \draw[line width=0.1cm] (0,0) -- (0,-.4);
        \draw[line width=0.1cm] (0,0) -- (1,0);
        \draw[doubline] (1,0) -- (1,-0.4);
        \draw[line width=0.1cm] (1,0) -- (1.5,0);
        \node[H] at (0,0) {};
        \node[B, T] at (1,0) {};
        \node[] at (2,0) {$+$};
        \pic{projector};
      \end{scope}

      \begin{scope}[shift={(8.25,0)}]
        \draw[line width=0.1cm] (0,0) -- (0,-.4);
        \draw[line width=0.1cm] (0,0) -- (1,0);
        \draw[doubline] (1,0) -- (1,-0.4);
        \draw[line width=0.1cm] (1,0) -- (1.5,0);
        \node[B] at (0,0) {};
        \node[H, T] at (1,0) {};
        \node[] at (2,0) {$+$};

        \pic{projector};
      \end{scope}

      \begin{scope}[shift={(11.25,0)}]
        \draw[line width=0.1cm] (0,0) -- (0,-.4);
        \draw[line width=0.1cm] (0,0) -- (1,0);
        \draw[doubline] (1,0) -- (1,-0.4);
        \draw[line width=0.1cm] (1,0) -- (1.5,0);
        \node[N] at (0,0) {};
        \node[HB, T] at (1,0) {};
        \node[] at (2,0) {$+$};

        \pic{projector};
      \end{scope}

      \begin{scope}[shift={(2.15,-1.75)}]
        \draw[line width=0.1cm] (0,0) -- (0,-.4);
        \draw[line width=0.1cm] (0,0) -- (1,0);
        \draw[doubline] (1,0) -- (1,-0.4);
        \draw[line width=0.1cm] (1,0) -- (1.75,0);
        \node[HB, Ch] at (0.55,0) {};

        \pic{projector};
      \end{scope}
    \end{scope} 

    \ExtractCoordinate{scope1.south}
    \pgfmathparse{\YCoord-30}
    \edef\YCoord{\pgfmathresult}
    \begin{scope}[shift={(0,-3.5)}, local bounding box=scope2]
      \node[anchor=west] at (-1.8,0) {$EBCh_1'$};

      \begin{scope}[xshift=-0.1cm]
        \draw[doubline] (1,0) -- (1,-0.75);
        \draw[line width=0.1cm] (0,0) -- (0,-0.75);
        \draw[line width=0.1cm] (1,0) -- (1.75,0);
        \node[HB, Ch] at (0.55,0) {};
        \node[] at (2.25,0) {$=$};
      \end{scope}

      \begin{scope}[shift={(3.25,0)}]
        \draw[line width=0.1cm] (0,0) -- (0,-0.4);
        \draw[line width=0.1cm] (0,0) -- (1,0);
        \draw[doubline] (1,0) -- (1,-0.4);
        \draw[doubline] (2,0) -- (2,-0.75);
        \draw[line width=0.1cm] (2,0) -- (2.75,0);

        \node[N] at (0,0) {};
        \node[HB, TTh] at (1.5,0) {};
        \node[] at (3.25,0) {$+$};

        \pic{projector};
      \end{scope}

      \begin{scope}[shift={(7.5,0)}]
        \draw[line width=0.1cm] (0,0) -- (0,-0.4);
        \draw[line width=0.1cm] (0,0) -- (1,0);
        \draw[doubline] (1,0) -- (1,-0.4);
        \draw[doubline] (2,0) -- (2,-0.75);
        \draw[line width=0.1cm] (2,0) -- (2.75,0);

        \node[B] at (0,0) {};
        \node[H, TTh] at (1.5,0) {};

        \pic{projector};
      \end{scope}

    \end{scope} 

    \ExtractCoordinate{scope2.south}
    \pgfmathparse{\YCoord-50}
    \edef\YCoord{\pgfmathresult}
    \begin{scope}[shift={(0,-6)}, local bounding box=scope3]
      \node[anchor=west] at (-1.8,0) {$EBT_4'$};

      \begin{scope}
        \draw[line width=0.1cm] (0,0) -- (0,0.5);
        \draw[doubline] (0,0) -- (0.75,0);
        \draw[line width=0.1cm] (0,0) -- (0,-0.5);
        \node[HB, Th] at (0,0) {};
        \node[] at (1.25,0) {$=$};
      \end{scope}

      \begin{scope}[shift={(2.25,0)}]
        \node[Hconnecth] at (0.5,0) {};

        \draw[line width=0.1cm] (0,0) -- (0,0.4);
        \draw[doubline] (0,0) -- (1,0);
        \draw[line width=0.1cm] (0,0) -- (0,-0.4);

        \draw[doubline] (1,0) -- (1,0.4);
        \draw[doubline] (1,0) -- (1.5,0);
        \draw[doubline] (1,0) -- (1,-0.4);

        \node[B, Th] at (0,0) {};
        \node[tensorxin] at (0.15,0) {$\times$};
        \node[tensorx] at (1,0) {$\times$};
        \node[] at (2,0) {$+$};

        \pic{projector};
        \pic{projector up};
      \end{scope}

      \begin{scope}[shift={(5.25,0)}]
        \node[Hconnecth] at (0.5,0) {};

        \draw[line width=0.1cm] (0,0) -- (0,0.4);
        \draw[doubline] (0,0) -- (1,0);
        \draw[line width=0.1cm] (0,0) -- (0,-0.4);

        \draw[doubline] (1,0) -- (1,0.4);
        \draw[doubline] (1,0) -- (1.5,0);
        \draw[doubline] (1,0) -- (1,-0.4);

        \node[N, Th] at (0,0) {};
        \node[tensorxin] at (0.15,0) {$\times$};
        \node[tensorxb] at (1,0) {$\times$};
        \node[] at (2,0) {$+$};

        \pic{projector};
        \pic{projector up};
      \end{scope}

      \begin{scope}[shift={(8.25,0)}]
        \draw[line width=0.1cm] (0,0) -- (0,0.4);
        \draw[doubline] (0,0) -- (1,0);
        \draw[line width=0.1cm] (0,0) -- (0,-0.4);

        \draw[doubline] (1,0) -- (1,0.4);
        \draw[doubline] (1,0) -- (1.5,0);
        \draw[doubline] (1,0) -- (1,-0.4);

        \node[HB, Th] at (0,0) {};
        \node[tensor] at (1,0) {};
        \node[] at (2,0) {$+$};

        \pic{projector};
        \pic{projector up};
      \end{scope}

      \begin{scope}[shift={(11.25,0)}]
        \draw[line width=0.1cm] (0,0) -- (0,0.4);
        \draw[doubline] (0,0) -- (1,0);
        \draw[line width=0.1cm] (0,0) -- (0,-0.4);

        \draw[doubline] (1,0) -- (1,0.4);
        \draw[doubline] (1,0) -- (1.5,0);
        \draw[doubline] (1,0) -- (1,-0.4);

        \node[H, Th] at (0,0) {};
        \node[tensorb] at (1,0) {};

        \pic{projector};
        \pic{projector up};
      \end{scope}

    \end{scope} 

    \ExtractCoordinate{scope3.south}
    \pgfmathparse{\YCoord-70}
    \edef\YCoord{\pgfmathresult}
    \begin{scope}[shift={(0,-9)}, local bounding box=scope4]
      \node[anchor=west] at (-1.8,0.5) {$EBCv_1'$};

      \begin{scope}

        \draw[line width=0.1cm] (0,1) -- (0.75,1);
        \draw[doubline] (0,0) -- (0.75,0);
        \draw[line width=0.1cm] (0,0) -- (0,-0.5);

        \node[HB, Cv] at (0,0.55) {};
        \node[] at (1.25,0.5) {$=$};

      \end{scope}

      \begin{scope}[shift={(2.25,0)}]
        \node[Hconnectv] at (1,0.5) {};

        \draw[line width=0.1cm] (0,1) -- (1,1);
        \draw[line width=0.1cm] (0,0) -- (0,1);
        \draw[line width=0.1cm] (0,0) -- (0,-0.4);
        \draw[line width=0.1cm] (1,1) -- (1.5,1);
        \draw[doubline] (0,0) -- (1,0);
        \draw[doubline] (1,0) -- (1,1);
        \draw[doubline] (1,0) -- (1,-0.4);
        \draw[doubline] (1,0) -- (1.5,0);

        \node[N] at (0,1) {};
        \node[N,Th] at (0,0) {};
        \node[B,T] at (1,1) {};
        \node[tensorxin] at (1,0.85) {$\times$};
        \node[tensorx] at (1,0) {$\times$};
        \node[] at (2,0.5) {$+$};

        \pic{projector};
      \end{scope}

      \begin{scope}[shift={(5.25,0)}]
        \node[Hconnectv] at (1,0.5) {};

        \draw[line width=0.1cm] (0,1) -- (1,1);
        \draw[line width=0.1cm] (0,0) -- (0,1);
        \draw[line width=0.1cm] (0,0) -- (0,-0.4);
        \draw[line width=0.1cm] (1,1) -- (1.5,1);
        \draw[doubline] (0,0) -- (1,0);
        \draw[doubline] (1,0) -- (1,1);
        \draw[doubline] (1,0) -- (1,-0.4);
        \draw[doubline] (1,0) -- (1.5,0);

        \node[N] at (0,1) {};
        \node[N,Th] at (0,0) {};
        \node[N,T] at (1,1) {};
        \node[tensorxin] at (1,0.85) {$\times$};
        \node[tensorxb] at (1,0) {$\times$};
        \node[] at (2,0.5) {$+$};

        \pic{projector};
      \end{scope}

      \begin{scope}[shift={(8.25,0)}]
        \node[Hconnectv] at (1,0.5) {};

        \draw[line width=0.1cm] (0,1) -- (1,1);
        \draw[line width=0.1cm] (0,0) -- (0,1);
        \draw[line width=0.1cm] (0,0) -- (0,-0.4);
        \draw[line width=0.1cm] (1,1) -- (1.5,1);
        \draw[doubline] (0,0) -- (1,0);
        \draw[doubline] (1,0) -- (1,1);
        \draw[doubline] (1,0) -- (1,-0.4);
        \draw[doubline] (1,0) -- (1.5,0);

        \node[B] at (0,1) {};
        \node[N,Th] at (0,0) {};
        \node[N,T] at (1,1) {};
        \node[tensorxin] at (1,0.85) {$\times$};
        \node[tensorx] at (1,0) {$\times$};
        \node[] at (2,0.5) {$+$};

        \pic{projector};
      \end{scope}

      \begin{scope}[shift={(11.25,0)}]
        \node[Hconnectv] at (1,0.5) {};

        \draw[line width=0.1cm] (0,1) -- (1,1);
        \draw[line width=0.1cm] (0,0) -- (0,1);
        \draw[line width=0.1cm] (0,0) -- (0,-0.4);
        \draw[line width=0.1cm] (1,1) -- (1.5,1);
        \draw[doubline] (0,0) -- (1,0);
        \draw[doubline] (1,0) -- (1,1);
        \draw[doubline] (1,0) -- (1,-0.4);
        \draw[doubline] (1,0) -- (1.5,0);

        \node[N] at (0,1) {};
        \node[B,Th] at (0,0) {};
        \node[N,T] at (1,1) {};
        \node[tensorxin] at (1,0.85) {$\times$};
        \node[tensorx] at (1,0) {$\times$};
        \node[] at (2,0.5) {$+$};

        \pic{projector};
      \end{scope}

      \begin{scope}[shift={(2.25,-2.75)}]
        \draw[line width=0.1cm] (0,1) -- (1,1);
        \draw[line width=0.1cm] (0,0) -- (0,-0.4);
        \draw[line width=0.1cm] (1,1) -- (1.5,1);
        \draw[doubline] (0,0) -- (1,0);
        \draw[doubline] (1,0) -- (1,1);
        \draw[doubline] (1,0) -- (1,-0.4);
        \draw[doubline] (1,0) -- (1.5,0);

        \node[HB, Cv] at (0,0.55) {};
        \node[N,T] at (1,1) {};
        \node[tensor] at (1,0) {};
        \node[] at (2,0.5) {$+$};

        \pic{projector};
      \end{scope}

      \begin{scope}[shift={(5.25,-2.75)}]
        \draw[line width=0.1cm] (0,1) -- (1,1);
        \draw[line width=0.1cm] (0,0) -- (0,-0.4);
        \draw[line width=0.1cm] (1,1) -- (1.5,1);
        \draw[doubline] (0,0) -- (1,0);
        \draw[doubline] (1,0) -- (1,1);
        \draw[doubline] (1,0) -- (1,-0.4);
        \draw[doubline] (1,0) -- (1.5,0);

        \node[H, Cv] at (0,0.55) {};
        \node[N,T] at (1,1) {};
        \node[tensorb] at (1,0) {};
        \node[] at (2,0.5) {$+$};

        \pic{projector};
      \end{scope}

      \begin{scope}[shift={(8.25,-2.75)}]
        \draw[line width=0.1cm] (0,1) -- (1,1);
        \draw[line width=0.1cm] (0,0) -- (0,-0.4);
        \draw[line width=0.1cm] (1,1) -- (1.5,1);
        \draw[doubline] (0,0) -- (1,0);
        \draw[doubline] (1,0) -- (1,1);
        \draw[doubline] (1,0) -- (1,-0.4);
        \draw[doubline] (1,0) -- (1.5,0);

        \node[H, Cv] at (0,0.55) {};
        \node[B,T] at (1,1) {};
        \node[tensor] at (1,0) {};

        \pic{projector};
        \node[] at (0,-1.5) {};
      \end{scope}
    \end{scope} 

    \ExtractCoordinate{scope4.south}
    \pgfmathparse{\YCoord-90}
    \edef\YCoord{\pgfmathresult}
    \begin{scope}[shift={(0,-15.5)}, local bounding box=scope5]
      \node[anchor=west] at (-1.8,0.5) {$EBTT_4'$};

      \begin{scope}
        \draw[line width=0.1cm] (0,1) -- (0,1.5);
        \draw[doubline] (0,1) -- (0.75,1);
        \draw[doubline] (0,0) -- (0.75,0);
        \draw[line width=0.1cm] (0,0) -- (0,-0.5);

        \node[HB, TTv] at (0,0.5) {};
        \node[] at (1.25,0.5) {$=$};

      \end{scope}

      \begin{scope}[shift={(2.25,0)}]
        \node[Hconnectv] at (1,0.5) {};

        \draw[line width=0.1cm] (0,0) -- (0,1);
        \draw[line width=0.1cm] (0,1) -- (0,1.4);
        \draw[doubline] (0,1) -- (1,1);
        \draw[doubline] (0,0) -- (1,0);
        \draw[line width=0.1cm] (0,0) -- (0,-0.4);
        \draw[doubline] (1,1) -- (1,1.4);
        \draw[doubline] (1,1) -- (1.5,1);
        \draw[doubline] (1,1) -- (1,0);
        \draw[doubline] (1,0) -- (1.5,0);
        \draw[doubline] (1,0) -- (1,-0.4);

        \node[N, Th] at (0,0) {};
        \node[B, Th] at (0,1) {};
        \node[tensorx] at (1,1) {$\times$};
        \node[tensorx] at (1,0) {$\times$};
        \node[] at (2,0.5) {$+$};

        \pic{projector};
        \pic{projector up high};
      \end{scope}

      \begin{scope}[shift={(5.25,0)}]
        \node[Hconnectv] at (1,0.5) {};

        \draw[line width=0.1cm] (0,0) -- (0,1);
        \draw[line width=0.1cm] (0,1) -- (0,1.4);
        \draw[doubline] (0,1) -- (1,1);
        \draw[doubline] (0,0) -- (1,0);
        \draw[line width=0.1cm] (0,0) -- (0,-0.4);
        \draw[doubline] (1,1) -- (1,1.4);
        \draw[doubline] (1,1) -- (1.5,1);
        \draw[doubline] (1,1) -- (1,0);
        \draw[doubline] (1,0) -- (1.4,0);
        \draw[doubline] (1,0) -- (1,-0.5);

        \node[B, Th] at (0,0) {};
        \node[N, Th] at (0,1) {};
        \node[tensorx] at (1,1) {$\times$};
        \node[tensorx] at (1,0) {$\times$};
        \node[] at (2,0.5) {$+$};

        \if\phases1
          \node[] at (1.65,1.75) {$\cdot~e^{i k_y}$};
        \fi

        \pic{projector};
        \pic{projector up high};

      \end{scope}

      \begin{scope}[shift={(8.25,0)}]
        \node[Hconnectv] at (1,0.5) {};

        \draw[line width=0.1cm] (0,0) -- (0,1);
        \draw[line width=0.1cm] (0,1) -- (0,1.4);
        \draw[doubline] (0,1) -- (1,1);
        \draw[doubline] (0,0) -- (1,0);
        \draw[line width=0.1cm] (0,0) -- (0,-0.4);
        \draw[doubline] (1,1) -- (1,1.4);
        \draw[doubline] (1,1) -- (1.5,1);
        \draw[doubline] (1,1) -- (1,0);
        \draw[doubline] (1,0) -- (1.5,0);
        \draw[doubline] (1,0) -- (1,-0.4);

        \node[N, Th] at (0,0) {};
        \node[N, Th] at (0,1) {};
        \node[tensorxb] at (1,1) {$\times$};
        \node[tensorx] at (1,0) {$\times$};
        \node[] at (2,0.5) {$+$};

        \pic{projector};
        \pic{projector up high};
      \end{scope}

      \begin{scope}[shift={(11.25,0)}]
        \node[Hconnectv] at (1,0.5) {};

        \draw[line width=0.1cm] (0,0) -- (0,1);
        \draw[line width=0.1cm] (0,1) -- (0,1.4);
        \draw[doubline] (0,1) -- (1,1);
        \draw[doubline] (0,0) -- (1,0);
        \draw[line width=0.1cm] (0,0) -- (0,-0.4);
        \draw[doubline] (1,1) -- (1,1.4);
        \draw[doubline] (1,1) -- (1.5,1);
        \draw[doubline] (1,1) -- (1,0);
        \draw[doubline] (1,0) -- (1.5,0);
        \draw[doubline] (1,0) -- (1,-0.4);

        \node[N, Th] at (0,0) {};
        \node[N, Th] at (0,1) {};
        \node[tensorx] at (1,1) {$\times$};
        \node[tensorxb] at (1,0) {$\times$};
        \node[] at (2,0.5) {$+$};

        \if\phases1
          \node[] at (1.65,1.75) {$\cdot~e^{i k_y}$};
        \fi

        \pic{projector};
        \pic{projector up high};
      \end{scope}

      \begin{scope}[shift={(2.25,-3.5)}]
        \draw[line width=0.1cm] (0,1) -- (0,1.4);
        \draw[doubline] (0,1) -- (1,1);
        \draw[doubline] (0,0) -- (1,0);
        \draw[line width=0.1cm] (0,0) -- (0,-0.4);
        \draw[doubline] (1,1) -- (1,1.4);
        \draw[doubline] (1,1) -- (1.5,1);
        \draw[doubline] (1,1) -- (1,0);
        \draw[doubline] (1,0) -- (1.5,0);
        \draw[doubline] (1,0) -- (1,-0.4);

        \node[HB, TTv] at (0,0.5) {};
        \node[tensor] at (1,1) {};
        \node[tensor] at (1,0) {};
        \node[] at (2,0.5) {$+$};

        \pic{projector};
        \pic{projector up high};
      \end{scope}

      \begin{scope}[shift={(5.25,-3.5)}]
        \draw[line width=0.1cm] (0,1) -- (0,1.4);
        \draw[doubline] (0,1) -- (1,1);
        \draw[doubline] (0,0) -- (1,0);
        \draw[line width=0.1cm] (0,0) -- (0,-0.4);
        \draw[doubline] (1,1) -- (1,1.4);
        \draw[doubline] (1,1) -- (1.5,1);
        \draw[doubline] (1,1) -- (1,0);
        \draw[doubline] (1,0) -- (1.5,0);
        \draw[doubline] (1,0) -- (1,-0.4);

        \node[H, TTv] at (0,0.5) {};
        \node[tensorb] at (1,1) {};
        \node[tensor] at (1,0) {};
        \node[] at (2,0.5) {$+$};

        \pic{projector};
        \pic{projector up high};
      \end{scope}

      \begin{scope}[shift={(8.25,-3.5)}]
        \draw[line width=0.1cm] (0,1) -- (0,1.4);
        \draw[doubline] (0,1) -- (1,1);
        \draw[doubline] (0,0) -- (1,0);
        \draw[line width=0.1cm] (0,0) -- (0,-0.4);
        \draw[doubline] (1,1) -- (1,1.4);
        \draw[doubline] (1,1) -- (1.5,1);
        \draw[doubline] (1,1) -- (1,0);
        \draw[doubline] (1,0) -- (1.5,0);
        \draw[doubline] (1,0) -- (1,-0.4);

        \node[H, TTv] at (0,0.5) {};
        \node[tensor] at (1,1) {};
        \node[tensorb] at (1,0) {};

        \if\phases1
          \node[] at (1.65,1.75) {$\cdot~e^{i k_y}$};
        \fi

        \pic{projector};
        \pic{projector up high};
      \end{scope}
    \end{scope} 

    \ExtractCoordinate{scope5.south}
    \pgfmathparse{\YCoord-85}
    \edef\YCoord{\pgfmathresult}
    \begin{scope}[shift={(0,-21.5)}, local bounding box=scope6]
      \node[anchor=west] at (-1.8,0) {$EBT_4o'$};

      \begin{scope}
        \draw[line width=0.1cm] (0,0) -- (0,0.5);
        \draw[doubline] (0,0) -- (0.75,0);
        \draw[line width=0.1cm] (0,0) -- (0,-0.5);
        \node[B, Th] at (0,0) {};
        \node[tensorxin] at (0.15,0) {$\times$};
        \node[] at (1.25,0) {$=$};
      \end{scope}

      \begin{scope}[shift={(2.25,0)}]
        \draw[line width=0.1cm] (0,0) -- (0,0.4);
        \draw[doubline] (0,0) -- (1,0);
        \draw[line width=0.1cm] (0,0) -- (0,-0.4);

        \draw[doubline] (1,0) -- (1,0.4);
        \draw[doubline] (1,0) -- (1.5,0);
        \draw[doubline] (1,0) -- (1,-0.4);

        \node[N, Th] at (0,0) {};
        \node[tensorxb] at (1,0) {$\times$};
        \node[] at (2,0) {$+$};

        \pic{projector};
        \pic{projector up};
      \end{scope}

      \begin{scope}[shift={(5.25,0)}]
        \draw[line width=0.1cm] (0,0) -- (0,0.4);
        \draw[doubline] (0,0) -- (1,0);
        \draw[line width=0.1cm] (0,0) -- (0,-0.4);

        \draw[doubline] (1,0) -- (1,0.4);
        \draw[doubline] (1,0) -- (1.5,0);
        \draw[doubline] (1,0) -- (1,-0.4);

        \node[B, Th] at (0,0) {};
        \node[tensorx] at (1,0) {$\times$};

        \pic{projector};
        \pic{projector up};
      \end{scope}

    \end{scope}
  \end{tikzpicture}
    \caption{All tensors necessary for one \emph{left move} of the combined CTM scheme, where the boundary tensors include all terms with a combination of a $B$ tensor and a Hamiltonian within the regions they represent.}
    \label{fig:main_tensors}
  \end{figure*}
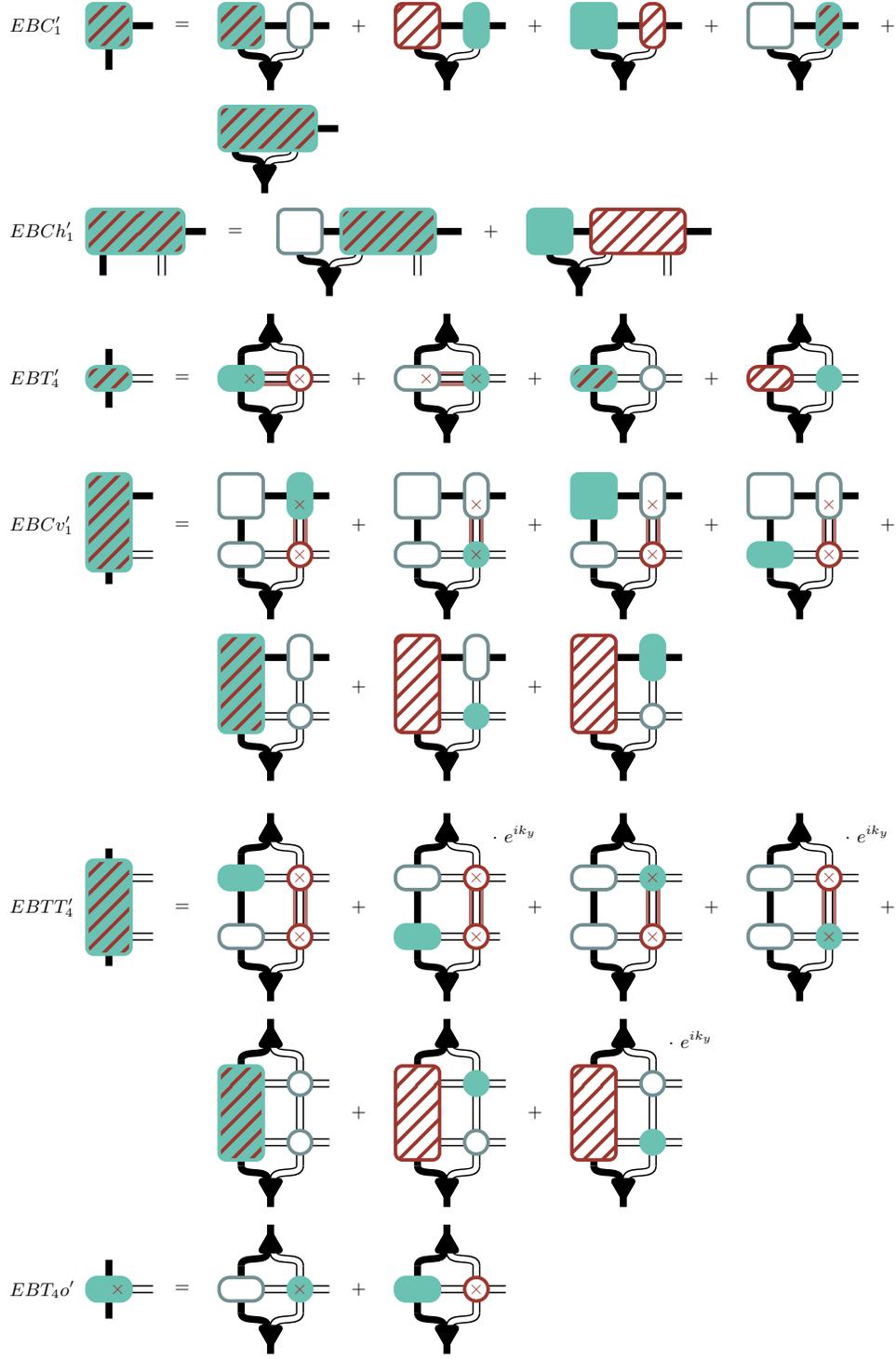

It would be interesting to apply these methods to other models where a description of the low-energy physics in terms of collective excitations may be valid.
The growth of the basis size for the overlap matrices with the bond dimension remains challenging.
However, since the main computation involves evaluating matrix elements separately, the algorithm can be run in parallel for large scale computations, or the eigenvalue problem can be solved iteratively for the few lowest eigenvalues.
Judicious preselection of relevant basis vectors would likely be greatly beneficial in reducing this cost, as well as more insight in the dependence on the gauge of the ground-state iPEPS.

\begin{acknowledgments}
We acknowledge many helpful discussions with L.~Vanderstraeten. This project has received funding from the European Research Council (ERC) under the European Union's Horizon 2020 research and innovation programme (grant agreement No 677061). This work is part of the D-ITP consortium, a program of the Netherlands Organization for Scientific Research (NWO) that is funded by the Dutch Ministry of Education, Culture and Science~(OCW). 
\end{acknowledgments}


\appendix
\section{Contraction scheme}
\label{sec:contraction_scheme}

In this appendix we show the contractions that are required for the boundary tensors that include contributions from both $B$ tensors and Hamiltonian terms in the same sector, in diagrammatic notation.
We limit the diagrams only to those relevant to an absorption of a column of sites to the left side, as in the discussion in~\Cref{ssub:main_scheme}.

The definition of the various shapes and symbols can be found in~\Cref{ssub:main_scheme}; the meaning of the coloring and shading is shown in~\Cref{fig:diag_colors}.
An update step of all boundary tensors on the left side consists of the contractions shown in~\Cref{fig:main_tensors}.
The last two types of boundary tensors at the bottom of the figure play a role in the contractions on the previous page, but do not themselves appear in any computation of expectation values.

\bibliographystyle{apsrev4-1}
\bibliography{refs}

\end{document}